\documentclass[10pt,journal,compsoc]{IEEEtran}
\ifCLASSOPTIONcompsoc
  \usepackage[nocompress]{cite}
\else
  \usepackage{cite}
\fi
\ifCLASSINFOpdf
  \usepackage[pdftex]{graphicx}
  \DeclareGraphicsExtensions{.pdf,.jpeg,.png,.jpg}
\else
  \usepackage[dvips]{graphicx}

\fi
\ifCLASSOPTIONcompsoc
  \usepackage[caption=false,font=footnotesize,labelfont=sf,textfont=sf]{subfig}
\else
  \usepackage[caption=false,font=footnotesize]{subfig}
\fi
\usepackage{balance}
\usepackage{pbalance}
\usepackage{url}
\usepackage{hyperref}
\usepackage{enumerate} % More powerful enumeration options
\usepackage[numbers, sort]{natbib} % Better looking bibliographies
\usepackage{flushend} % Automatically balance columns at the end of the paper
\usepackage{booktabs} % For modern, less cluttered tables
\usepackage{color} % For access to colours
\usepackage{pbalance}
\usepackage{tcolorbox}
\usepackage{orcidlink}
\usepackage[numbers, sort]{natbib} % Better looking bibliographies
\newcounter{finding_counter}
\newcommand{\mhmd}[1]{\textcolor{red}{#1}}

\usepackage{listings}
\usepackage{hyperref}
\usepackage{enumitem}
\usepackage{fancyhdr} 
\usepackage{tikz}
\usepackage{pgfplots}
\usepackage{caption}
\usepackage{subcaption}
\usepackage{algpseudocode}
\usepackage{algorithm}
\usepackage{multirow}
\usepackage{booktabs}
\usepackage{tabularx}
\usepackage{comment}
\usepackage{url}
\usepackage[flushleft]{threeparttable}
\usepackage[export]{adjustbox}[2011/08/13]

\usepackage{pdflscape}
\usepackage{afterpage}
\usepackage{rotating,graphicx}

\usepackage{makecell}
	
\usepackage[switch]{lineno}

\usepackage{listings}
\usepackage{color}
\usepackage[numbers, sort]{natbib} % Better looking bibliographies

\definecolor{codegreen}{rgb}{0,0.6,0}
\definecolor{codegray}{rgb}{0.5,0.5,0.5}
\definecolor{codepurple}{rgb}{0.58,0,0.82}
\definecolor{backcolour}{rgb}{0.95,0.95,0.92}
\definecolor{dkgreen}{rgb}{0,0.6,0}
\definecolor{gray}{rgb}{0.5,0.5,0.5}
\definecolor{mauve}{rgb}{0.58,0,0.82}

\lstdefinestyle{mystyle}{
     backgroundcolor=\color{backcolour},   
    commentstyle=\color{codegreen},
    keywordstyle=\color{magenta},
    numberstyle=\tiny\color{codegray},
    stringstyle=\color{codepurple},
    basicstyle=\ttfamily\footnotesize,
    breakatwhitespace=false,         
    breaklines=true,                 
    captionpos=b,                    
    keepspaces=true,                 
    numbers=left,                    
    numbersep=5pt,                  
    showspaces=false,                
    showstringspaces=false,
    showtabs=false,                  
    tabsize=2
}

\usepackage{listings}
\usepackage{xcolor}
\def\checkmark{\tikz\fill[scale=0.23](0,.35) -- (.25,0) -- (1,.7) -- (.25,.15) -- cycle;}
\begin{document}
%
% paper title
% Titles are generally capitalized except for words such as a, an, and, as,
% at, but, by, for, in, nor, of, on, or, the, to and up, which are usually
% not capitalized unless they are the first or last word of the title.
% Linebreaks \\ can be used within to get better formatting as desired.
% Do not put math or special symbols in the title.
\title{AddressWatcher: Sanitizer-Based Localization of Memory Leak Fixes}
\IEEEtitleabstractindextext{%
\begin{abstract}
Memory leak bugs are a major problem in C/C++ programs. They occur when memory objects are not deallocated.
Developers need to manually deallocate these objects to prevent memory leaks.
As such, several techniques have been proposed to automatically fix memory leaks.
Although proposed approaches have merit in automatically fixing memory leaks, they present limitations. 
Static-based approaches attempt to trace the complete semantics of memory object across all paths. However, they have scalability-related challenges when the target program has a large number of leaked paths.
On the other hand, dynamic approaches can spell out precise semantics of memory object only on a single execution path (not considering multiple execution paths).

In this paper, we complement prior approaches by designing and implementing a novel framework named \textit{AddressWatcher}.
AddressWatcher allows the semantics of a memory object to be tracked on multiple execution paths as a dynamic approach. 
Addresswatcher accomplishes this by using a leak database that is designed to allow storing and comparing different execution paths of a leak over several test cases.
% Also, AddressWatcher performs lightweight instrumentation during compile time that is utilized during the program execution to watch and track memory leak read/writes.
% For each leak, AddressWatcher tracks possible execution paths that a leaked object passes through and collects all stack traces where the leaked object is used.
% Finally, a stack that involves the deepest point in the program flow is suggested as a potential fix location for the leak.
%The framework utilizes these tools to perform a compile time instrumentation to watch and track memory leaks.
%In addition, AddressWatcher relies upon test suite coverage, which overcome the path explosion problem.
We conduct an evaluation of AddressWatcher over five popular open-source packages, namely binutils, openssh, tmux, openssl and git.
In 23 out of the 50 examined real-world memory leak bugs, AddressWatcher correctly points to a free location to fix memory leaks.
Moreover, to demonstrate the real-world impact of AddressWatcher, we submitted 25 pull requests (PRs) to 12 popular open-source project repositories. 
These PRs targeted the resolution of memory leaks within these repositories.
Among these, 21 PRs were merged, addressing 5 open GitHub issues. In fact, a critical fix prompted a new version release for the calc repository, a program used to find large primes.
% AddressWatcher automatically patched a gradual memory leak buildup in calc that triggered crashes when discovering larger primes over years of execution. 
Furthermore, our contributions through these PRs sparked intense discussions and appreciation in various repositories such as coturn, h2o, and radare2, highlighting the significant impact of AddressWatcher.
% Finally, our evaluation shows that AddressWatcher has a low performance overhead, i.e., 2.42-2.78 times over the base program.
% We compare our framework with Memfix, the state-of-the-art approach for fixing memory leaks, and find that Memfix fixes 20 bugs out of the examined 50 bugs.
% Furthermore, our analysis demonstrates that AddressWatcher can fix 11 memory leaks that Memfix cannot fix. 
% Finally, our evaluation shows that AddressWatcher has a performance overhead of 2.42-2.78 times over the base program.
% We develop a tool prototype for our framework (i.e., AddressWatcher) and make it publicly available.
\end{abstract}

% Note that keywords are not normally used for peerreview papers.
}

\author{
    \IEEEauthorblockN{Aniruddhan Murali\orcidlink{0000-0002-4405-1657}, \IEEEmembership{Student Member, IEEE}, Mahmoud Alfadel\orcidlink{0000-0002-2621-6104}, \IEEEmembership{Member, IEEE}, \\Meiyappan Nagappan\orcidlink{0000-0003-4533-4728}, Meng Xu\orcidlink{0009-0001-6364-4837} \IEEEmembership{Member, IEEE} and Chengnian Sun\orcidlink{0000-0002-0862-2491}, \IEEEmembership{Member, IEEE}~\thanks{All authors are with the David R. Cheriton School of Computer Science, University of Waterloo, Canada. 
    \\E-mail:\{a25murali, malfadel, mei.nagappan, meng.xu.cs, cnsun\}@uwaterloo.ca}}
}

% make the title area
\maketitle

% To allow for easy dual compilation without having to reenter the
% abstract/keywords data, the \IEEEtitleabstractindextext text will
% not be used in maketitle, but will appear (i.e., to be "transported")
% here as \IEEEdisplaynontitleabstractindextext when compsoc mode
% is not selected <OR> if conference mode is selected - because compsoc
% conference papers position the abstract like regular (non-compsoc)
% papers do!
\IEEEdisplaynontitleabstractindextext
% \IEEEdisplaynontitleabstractindextext has no effect when using
% compsoc under a non-conference mode.

% For peer review papers, you can put extra information on the cover
% page as needed:
% \ifCLASSOPTIONpeerreview
% \begin{center} \bfseries EDICS Category: 3-BBND \end{center}
% \fi
%
% For peerreview papers, this IEEEtran command inserts a page break and
% creates the second title. It will be ignored for other modes.
\IEEEpeerreviewmaketitle

\section{Introduction}
\label{sec:introduction}

% \IEEEPARstart{M}{emory} 
Memory leaks are common bugs in programming languages like C/C++.
They mainly occur when dynamically allocated objects are not deallocated.
Programming languages like C and C++ do not have automatic garbage collection, instead, they rely on developers to manually deallocate memory objects. Due to such a manual process, developers may forget to deallocate an object, causing a memory leak.

Memory leaks may have a large negative impact on software systems if not carefully examined and fixed.
In fact, memory leaks are direct sources of security vulnerabilities.
Attackers can utilize a memory leak to launch a denial of service (DoS) by crashing or hanging the program and taking advantage of other unexpected program behaviour resulting from low memory condition~\cite{b49}. 
Recently, a number of memory leak vulnerabilities have been disclosed in Linux kernel (e.g., CVE-2022-27819~\cite{b50}, CVE-2017-10810~\cite{b51}).
%For example, CVE-2022-27819 is a memory leak vulnerability in the Linux kernel. Through this bug, denial of service was possible through memory exhaustion when a large file is parsed by swhkd - a daemon used to easily add or remove hotkeys.
Such vulnerabilities in the kernel had severe consequences on system stability and availability~\cite{b28,b29,b30}.

%17 CVEs were assigned to Linux kernel memory leak bugs, 10 of these vulnerabilities were discovered in the past two years \textbf{\cite{b27}}.  Resource exhaustion vulnerabilities in the kernel has severe consequences on system stability and availability \textbf{\cite{b28},\cite{b29},\cite{b30}}.}

Fixing memory leaks manually is often time-consuming and error-prone for developers~\cite{b13}. 
Hence, prior work focused on designing and implementing techniques to address challenges of automatically fixing memory leak bugs~\cite{b13,b18,b19,b45,b12}.
Many of these prominent techniques leverage static and dynamic approaches to fixing memory leak bugs,
although very few of these
approaches have been open-sourced~\cite{b13,b18}
(both of them being static approaches).

An exemplary static analysis tool is Memfix~\cite{b13}. 
%It is a static analysis approach to fixing memory leaks, heap use-after-free and double frees at the same time.
Memfix identifies all paths involving an allocated memory. 
%Then, it inserts deallocation statements at the last-use points on all identified paths.
It models the problem of identifying a set of deallocation statements on these identified paths as an exact cover problem. 
It then uses a SAT solver 
%which is an algorithm to establish satisfiability of a given problem described in a Boolean logic 
to find the solution to the exact cover problem~\cite{b10}.
%Memfix fixes not only memory leaks, but also heap-use-after-free and double frees. 
The solution suggested by Memfix is always a safe fix. 
However, Memfix cannot resolve program paths in the presence of function pointers and recursion, and it errors out when there is an explosion of leaked program paths.

Furthermore, prior work has proposed dynamic analysis techniques to mitigate the problem of over-approximation by static approaches. 
LeakPoint is one example of such an approach, which is a dynamic analysis tool that performs taint propagation on leaked objects~\cite{b12}. 
It identifies last-use sites of leaked objects and suggests candidate sites for leak fixing.
One limitation of LeakPoint is that the fix is limited to considering a given execution path.
 
\textcolor{black}{In this paper, we complement prior approaches (e.g., static approaches like Memfix) by proposing a new open-source dynamic approach for memory leak fixing, called \textit{AddressWatcher}.
Our approach aims to automatically identify locations where a memory leak should be fixed.
% {suggest} locations in which a memory leak should be fixed.
AddressWatcher is an iterative process that is designed to refine its fixes over several test runs as more memory leak paths are uncovered. 
AddressWatcher achieves this by cross-linking the runtime behavior of the same memory object across a fleet of test cases.}
The conventional views of static and dynamic analysis are as follows:

% It differs from static and dynamic analysis techniques as follows:}
\begin{itemize}
    \item \textcolor{black}{Static analysis techniques can spell out the complete semantics of a memory object (from allocation to free) on all paths, but this is not precise (e.g., Memfix~\cite{b13}, Leakfix~\cite{b18}).}
    \item \textcolor{black}{Dynamic analysis techniques can spell out the precise semantics of a memory object on a particular execution path. This is enough for leak detection but less useful for suggesting a fix covering multiple execution paths (e.g., LeakPoint~\cite{b12}, LeakChaser~\cite{b52}).}
\end{itemize}

\textcolor{black}{AddressWatcher offers a way to combine the strengths of both views--it allows the semantics of a memory object to be tracked on multiple execution paths, only bounded by the quality of the test suite.
In particular, AddressWatcher uses a leak database as a dynamic analysis technique, storing and comparing execution traces of leaks over several test runs. 
It also complements prior static analysis tools by considering memory leak cases that are unable to be fixed by such static-based analysis, i.e., our approach relies upon test case runs to track execution paths of leaks, and hence, it does not suffer from issues related to ``path explosion'' problems that are present in static-based approaches.
Moreover, our approach addresses the problem of slowdown due to dynamic binary instrumentation in certain dynamic-based approaches (e.g., LeakPoint) by using sanitizer-based and light-weight compile-time instrumentation.}

\vspace{1mm}
\noindent
\textbf{Contributions.} The key contributions of this paper are as follows:

\begin{enumerate}
\item We present AddressWatcher, an automated dynamic analysis tool that suggests locations for memory leak fixes in C/C++ programs.
AddressWatcher introduces the concept of using shadow memory to tag memory and eventually suggests a bug-fix location. 
% Previous approaches instead used shadow memory mainly to detect memory errors~\cite{b53,b2,b41,b54}. 
AddressWatcher can suggest multiple free locations for a given leak after considering all relevant execution traces in a leak database.
% To our best knowledge,  AddressWatcher is the first to utilize sanitizer-based techniques (i.e., ASAN \& LSAN) to perform compile time instrumentation to track memory leak addresses. 
% \item AddressWatcher introduces the concept of using shadow memory to tag memory and eventually suggest bug fix location. Previous approaches instead used shadow memory mainly to detect memory errors~\cite{b53,b2,b41,b54}.

% \item Using shadow memory to tag and instrumentation to track memory regions that have been leaked in previous binary runs.
% \item A novel approach integrating LSAN and ASAN together in AddressWatcher with minimal changes

\item We examine the effectiveness of AddressWatcher on a set of 50 memory leak bugs in popular packages, followed by a qualitative analysis. 
% Additionally, we evaluate the efficiency of AddressWatcher in suggesting memory leak fixes. 
We compare the fix locations suggested by AddressWatcher with Memfix~\cite{b13}. 
% Our comparison provided a deeper understanding of which tools are more likely to succeed in fixing a given memory leak, taking into consideration several underlying factors.
\item
We demonstrate the practical relevance of AddressWatcher by submitting 25 PRs to major open-source projects with memory leak issues. Of these, 21 were approved and merged, resolving 5 open GitHub issues~\cite{iniparser-issue,h2o-issue,snmp-issue,whois-issue,nanonng-issue}.
% while 4 PRs are under review.
% Notably, AddressWatcher patched a gradual memory leak buildup in calc~\cite{calc-Release} that triggered crashes when discovering larger primes over years of execution. The fix was deemed so critical that it prompted a new version release for the calc repository. It also led to intense discussions and appreciation for our fixes in coturn, h2o and radare2 repositories highlighting the impact of the fixes~\cite{coturn-disc, h2o1, radare10}.
% \item \textcolor{black}{We submit 25 Pull Requests (PRs) fixing leaks across 12 diverse repositories using the suggestions from AdddressWatcher. Among these PRs, 21 were merged, and 5 are pending review. As a result of these merges, 5 open issues were resolved~\cite{iniparser-issue,h2o-issue,snmp-issue,whois-issue,nanonng-issue}. Additionally, our fixes led to a new version release for calc repository~\cite{calc-Release} and intense discussions on memory safety in coturn repository highlighting the impact of the fixes~\cite{coturn-disc}.}

% \item We conduct an evaluation of AddressWatcher over a set of memory leak bugs in popular packages, followed by a qualitative analysis. We also evaluate the efficiency of AddressWatcher in suggesting memory leak fixes.
% We also compare fixes suggested by AddressWatcher with the state-of-the-art approach for statically fixing memory leaks, Memfix~\cite{b13}. 
% Our comparison provides more understanding about which tools are more likely to succeed in fixing a given memory leak considering several underlying factors.

\item We develop a prototype tool for AddressWatcher. The tool and benchmarks are made publicly available~\cite{b25}.
% an open-source implementation of AW for C/C++ ~\cite{repo}.
\end{enumerate}

% AddressWatcher’s novelty lies in its ability to watch memory leaks in binary programs, enabling the identification of a potential fix location in an efficient manner. 
% Also, it uses special shadow memory to tag the stored leaked objects.
% We note that in previous work shadow memory was mainly used to \textit{detect} memory errors. AddressWatcher instead utilizes shadow memory for the purpose of bug fix suggestion.
% Finally, it introduces a new technique that utilizes instrumentation as checkpoints to enable profiling leak execution paths.

\section{Memory leaks \& sanitizers}\label{sec:Background}

In this section, we provide an overview of several concepts related to memory errors.

% that are utilized in this work and in other tools to detect memory errors.\\

\textbf{Shadow memory} is a duplicate region of memory used to mimic the state of actual process memory. It can be used to store any kind of information about the state of process memory~\cite{b53,b2,b41,b54}.
\textbf{Compile-time instrumentation} refers to the insertion of appropriate instructions into the application binary during compilation. These instructions can be used to detect the violation of a given property to identify memory errors~\cite{Serebryany2012,saeed2016tag}.
\textbf{Red zones} are fixed-size blocks of memory that are safe from modification by a given application. Red zones have been used to detect several memory errors. For example, a common technique to detect buffer overflows is to pad local and global variables with red zone buffers. If a read/write operation happens to a red zone then a buffer overflow has occurred. A common technique to identify the location of buffer overflow is to use compile-time instrumentation to insert checks at suspicious read and write locations. The instrumentation checks if the read/write happens to a red zone memory region. 
Several tools (e.g., ASAN~\cite{b2}, Dr. Memory~\cite{b53}) have been proposed to detect such scenarios by using shadow memory to encode the location of inserted red zone regions.
% , tools like ASAN ~\cite{b2} have utilized shadow memory to encode location of inserted redzone regions.

 %Red zones have also been used for the purpose of storing relevant information about the surrounding process memory. This can be useful to trace the source of certain memory errors. One such example is the case of memory leaks. 
 
 \textbf{Memory leak checking} is an example of a memory error whose source can be traced by red zones that store relevant information about the surrounding process memory.
 Whenever memory is allocated, a red zone region can store metadata such as a thread ID for allocating the memory, the size of allocated memory, and the program stack at which the memory was allocated. Tools such as LSAN~\cite{b5} ensure that when the allocated memory is freed, it is overwritten with a magic value. Before the program termination, LSAN detects memory leaks by identifying allocated memory surrounded by red zone regions that have not been overwritten with a magic value. The allocation stack is then retrieved from the red zone as the source of the memory leak.

\begin{lstlisting}[language=C++, caption=Leak detection vs leak fixing.
  ,label={fixvsdetect}]
     void read(int size) {
         char* p = (char*)malloc(size);
         fgets(p, size, stdin);
         if (*p == "\n") {
            // Path 1: Abrupt return
            // Leak p
            return;
         }
         // Path 2: Process p
         // Leak p
         return;
     }
\end{lstlisting}

 \textbf{Leak detection vs fixing.} \textcolor{black}{While memory leak detection and leak fixing are relevant, they require fundamentally different approaches to address them.
 % A dynamic leak detection technique such as LSAN~\cite{b5} checks for memory that is allocated but not freed only at program termination. It does not profile leaked memory because it is an additional overhead and is not required to detect a leak. 
 In the case of static leak detectors such as Saber~\cite{b1}, it is sufficient to track allocated memory being leaked along just} \textit{one} \textcolor{black}{path to confirm a leak. For example, in code listing~\ref{fixvsdetect}, user stdin input is stored in ``p". If input starts with a newline character, then we return abruptly on Path 1 (see line 7), which leaks allocated memory. Otherwise, we proceed to process user input normally on Path 2 (see line 11), which also leaks the memory. A static tool can analyze Path 1 alone to confirm the leak.  However, in order to fix the leak, a static technique must analyze}  \textit{all} \textcolor{black}{ paths that leak memory in order to fully deallocate the leak (by inserting frees along both Path 1 and Path 2).}

\section{Design}
\label{sec:design}
AddressWatcher is an approach that performs memory tracking to automatically suggest locations where a fix for a memory leak should be placed.

To put the problem more formally:
consider a memory allocation at code point $A$ within a program that is subsequently leaked. Assume that the allocated object is used at different code points $o_1, o_2, o_3 .... o_n$ within the program.
A use of an allocated object can refer to either a read or write to the allocated object.
Intuitively, the fix to this memory leak should be placed
after the last use of the allocated object.
Therefore, the core problem AddressWatcher aims to solve is
how to automatically identify the last use of a memory object.

% {\textcolor{red}{I don't think ``control-flow execution order'' is a standard and well-defined term.
% %
% As commented in the MS Team chat, I understand what you want to convey,
% but this ordering relation needs a much careful description otherwise
% it is an invitation for critics.
% %
% Here is what I have in mind,
% which hopefully can provide some food for thought,
% even in the presence of branches, loops, and recursions.
% %
% \begin{itemize}
%     \item When $o_i$ and $o_j$ reside in the same CFG,
%     $o_i >_s o_j$ iff $o_i$ \emph{post-dominates} $o_j$.
%     \item When $o_i$ and $o_j$ reside in different CFGs
%     (i.e., in different functions),
%     $o_i >_s o_j$ is defined by what you have in
%     Section Implementaiton-Comparing stacks
%     (BTW, that should not be in implementation,
%     it is a high-level construct which should belong here).
% \end{itemize}
% %
% I would also argue that,
% if we cannot establish the ordering relation between
% $o_i$ and $o_j$, AddressWatcher should find a new $o_k$
% which post-dominates both $o_i$ and $o_j$.
% %
% This $o_k$ should be the point where the fix needs to be placed.
% }}

On a high level, AddressWatcher relies on dynamic information
collected when executing test cases of the target program.
To illustrate, assuming
the testsuite for this program contains two different testcases $t_1$ and $t_2$. Let us assume that testcase $t_1$ executes a sequence of these code points $A, o_2, o_3, o_4$ etc. Similarly the other testcase $t2$ can execute a different sequence of code points $A, o_1, o_5$. AddressWatcher needs to identify that $o_4$ and $o_5$ are the last points where memory related to allocation $A$ was used and hence needs to be subsequently freed.

AddressWatcher breaks down the challenge of suggesting memory leak fixes into the following (D)esign components:

\begin{enumerate}
    \item[\textbf{(D1)}] \textbf{Instrument relevant code points.} Set up checkpoints at all relevant code points through binary instrumentation, i.e., $o_1,o_2,...o_n$.
    \item[\textbf{(D2)}] \textbf{Detect leak.} At the end of a given testcase, AddressWatcher must detect that the allocation at $A$ is not de-allocated.
    \item[\textbf{(D3)}] \textbf{Tag leak.} At a given allocation code point while executing a testcase, AddressWatcher must detect that the allocation was leaked on some other testcase in the past.
    \item[\textbf{(D4)}] \textbf{Track leak.} Identify when leaked memory is read/written at a given instrumented checkpoint.
    \item[\textbf{(D5)}] \textbf{Preserve leak and execution trace integrity across testcases.} Store leak information and execution trace of leaks in a database in a way that preserves integrity across multiple testcases.
    \item[\textbf{(D6)}] \textbf{Suggest fix location.} Compare all the different execution traces (eg. execution traces for testcases $t_1$ and $t_2$) to identify multiple last use points,
    after which a fix is suggested.
\end{enumerate}

% \noindent
% \textbf{The comparison operator.}
% We note that the code points $o_1, o_2, ... o_n$ are sorted by a comparison operator $>_s$ defined as follows:

% \begin{figure}[tb!]
% \includegraphics[width=0.8\linewidth]{images/Stackcomparison.png}
%     \caption{Example of comparison operator $>$s for comparing two stacks.}
%     \label{Stack comparison}
% \end{figure}

% \begin{itemize}
%     \item When $o_i$ and $o_j$ reside in the same function,
%     $o_i >_s o_j$ iff $o_i$ \emph{post-dominates} $o_j$,
%     i.e., $o_i$ is a code point after $o_j$ on all exit paths
%     in their enclosing function.
    
%     \item When $o_i$ and $o_j$ reside in different functions,
%     we consider their corresponding program stacks during execution labelled as S1 and S2. Consider an example case of S1 and S2 (Figure ~\ref{Stack comparison}).
%     We want to determine which stack represents a deeper point in the program according to the control flow logic.
%     That is, we take the first base layer of both stacks and compare the value 0x400 from S1 with the value 0x400 from S2. Since they are equal, we move to the next layer of the stacks, i.e., we compare 0x300 from S1 with 0x300 from S2. Once again, the values are equal, and hence, we move the comparison to the next layer. 
%     We can observe that 0x201 is greater than 0x200, meaning that 0x201 is a more latest point in the program. 
%     Hence, we arrive at the conclusion that S1 is greater than S2 according to this operator. Therefore $o_i >_s o_j$ 
% \end{itemize}

Each of the design components (mentioned above) represents unique challenges. 
In Section~\ref{sec:approachoverview}, we discuss how each of these challenges is handled.

\section{Proposed Approach}
\label{sec:approachoverview}

% AddressWatcher is an approach that performs memory address tracking to automatically suggest a location where a fix for a memory leak should be added. 

In the previous section, we discussed the design challenges that AddressWatcher aims to tackle. In this section, we discuss how each unique challenge is handled in the AddressWatcher framework.
% The approach modifies the existing infrastructure of  two tools, ASAN~\cite{b2} and LSAN~\cite{b5},
% (described in Section ~\ref{sec:Background}),
% with the aim of suggesting potential location for leak fixes. 
\graphicspath{ {./images/} }
\begin{figure*}[t!]
    \centering
    \includegraphics[width=\linewidth]{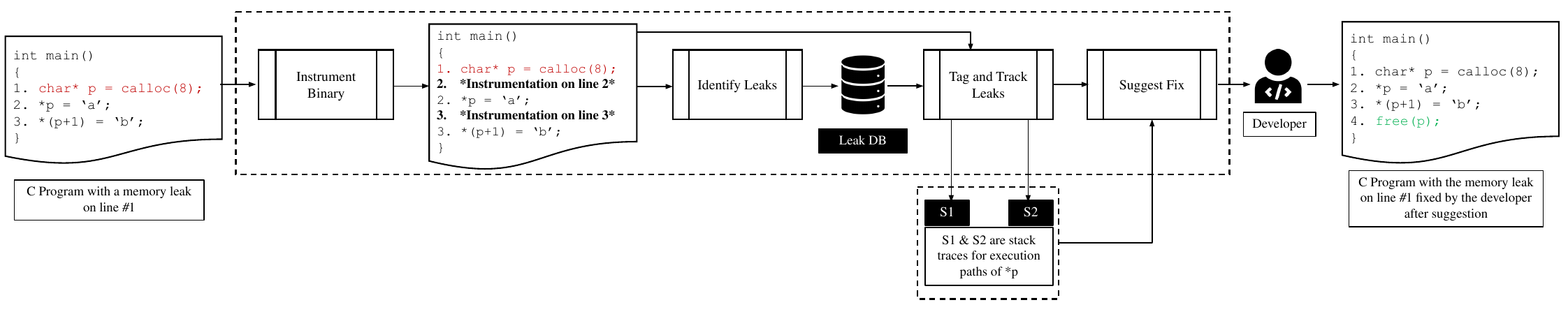}    \caption{An overview of our approach for suggesting a location of a memory leak fix.}
    %\mhmd{I think we need to add a stack S2 that presents the solution.}}
    \label{Approach}
\end{figure*} 
Figure \ref{Approach} shows an overview of our approach. The Figure shows a code program that suffers from a memory leak on line 1. Our goal is to suggest a location where a fix to the memory leak should be added, i.e., the approach aims to suggest a code line for adding a deallocation statement.
%In the given code example, a free statement should be added after line 3, in order to address the leak on line 1.
Our approach goes through four main steps to achieve the goal, (1) Adding a leak checker and instrumenting program binary; (2) Identifying memory leaks; (3) Tagging and tracking execution paths of memory leaks; (4) Suggesting fix location for memory leaks. These steps outline a concrete implementation to achieve each of the design components described in the previous section.

\iffalse
\begin{lstlisting}[language=C++, caption=Memory leak simple example\label{approach example}]
    int main {
        char *p = malloc(8);
        *p = 'a';
        *(p+1) = 'b';
        return 0;
    }
\end{lstlisting}
\fi

Next, we describe each step in detail using the code example shown in Figure \ref{Approach}. We explain how the approach goes through the four steps to suggest a fix location after line 3 for the memory leak that occurs in the code example.

\vspace{2mm}
\noindent
\textbf{Step 1: Adding a leak checker and instrumenting program binary.}
The first goal of this step is to obtain an instrumented binary of a given program. 
%To do so, we utilize code instrumentation provided by ASAN.
% We compile the program with ASAN code instrumentation.
The compiler adds AddressWatcher instrumentation before each read and write operation in the program.
%when executing the program under a test suite, ASAN \mhmd{during compile time, we run ASAN} adds new binary code before each read and write operation in the program.
As shown in Figure~\ref{Approach}, AddressWatcher adds some code before lines 2 and 3, since these lines contain read/write operations of program memory.
The instrumented binary is an essential artifact of the approach as it helps for tracking memory leaks (more on this in later steps). The AddressWatcher instrumentation behaves as a checkpoint while tracking reads and writes of memory objects. This enforces the design component (D1) described in the previous section.
%
% \textcolor{red}{We note that the AddressWatcher instrumentation is a modification of ASAN instrumentation~\cite{b2}. We discuss these modifications in section~\ref{sec:implementation}. }
%
% \textcolor{red}{(Don't say that we modify ASAN, instead,
% list the information you would like your instrumented logic to track.
% %
% To my understanding, what you need to track is:
% for each read and write, which memory object is touched.
% %
% This is actually a much lightweight requirement compared with
% what ASAN is doing.)}
The other goal of this instrumentation step is to
insert sanity-checking logic for
detecting memory leaks.
%
% so that AddressWatcher can search fixes for the detected leaks.
%
% {\textcolor{red}{Instead of saying we use LSAN,
% it is better to describe the leak detection logic.
% %
% I am not entirely familiar with the LSAN details,
% but the simplies (and perhaps most intuitive) form of
% leak detection algorithm is reference counting, i.e.,
% \begin{itemize}
%     \item Whenever a \texttt{malloc}-related call is invoked,
%     mark the returned pointer as allocated
%     \item Whenever a \texttt{free} is called,
%     mark the memory object as freed.
%     \item At the end of program execution,
%     whatever memory object is not freed is a leak.
% \end{itemize}
% }}
The output of this step is an instrumented binary of the given program.

\vspace{2mm}
\noindent
\textbf{Step 2: Identifying memory leaks.}
In the previous step, we are able to obtain an instrumented binary of the program. 
In this step, we use the instrumented binary and identify memory leak objects (e.g., code points that have malloc, calloc, and realloc allocations without certain deallocations in the program).
To achieve this step, 
%we utilize LSAN to detect stack traces of a leak object.
we execute the instrumented binary which essentially runs the existing test cases.
Just before the program is terminated while running the tests, the leak checker module (which was instrumented in step 1) is invoked to perform heap analysis to detect leaked objects (as described in Section~\ref{sec:Background}).
For the running example in Figure~\ref{Approach}, the leak checker identifies that there is a leaked object originating from the calloc function on line 1.
The leak checker detects leaks at the end of the given testcase which accomplishes the design step (D2).
In fact, the checker additionally stores the allocation stack traces of the leaks in a database, which we call a leak database. The allocation stack trace is the program stack when the leaked object is allocated.
Initially, the leak database is empty, however, after the first program execution, the leak database gets populated with the detected leaks.
This is one essential part of achieving the design component (D5). Subsequent testcases will then read from the leak database to match allocations with leaks from testcases in the past.
Such a database aims to help us for profiling and tracking the execution paths of each leaked object, as we describe in the next step.
%Note that our approach consists of multiple testcase runs. 
%In the next step, we perform execution path profiling of the detected leak.\\

\vspace{2mm}
\noindent
\textbf{Step 3: Tagging and tracking execution paths of memory leaks.}
So far we are able to only identify leaked objects of the program (stored in the leak database). In this step, our goal is to obtain the execution paths of the previously identified leaked objects. 
Obtaining the execution path of the identified leak is important to deduce the fix location in the next step.
The execution path of a leaked object is a list of stacks that represent code points that the leaked object passes through.
To obtain these stacks, we re-execute the program binary, and perform two tasks: 1) tagging leaks; and 2) {tracking} leaks.
% To do so, we need to navigate through the program one more time to track such paths.
% %(note that the first program run was performed in step 1).
% As such, we re-execute the program binary and utilize ASAN to help us with tracking the execution paths.

\textit{Tagging leaks.}
To enable tracking leaked objects during the program execution, we first need to distinguish leaked from non-leaked memory objects, i.e., we need to tag leaked objects to be tracked during the execution.
To perform tagging, we
% we utilize our modifications of ASAN. In fact, we modify a routine in ASAN library by adding a check point to 
examine if each allocation is a memory leak in a previous execution, by obtaining its allocation stack and then comparing it to the stacks stored in the leak database.
If there is a match with any stack in our leak database, we add a special value (i.e., a tag) to the leaks' shadow memory.
Given the code program in Figure~\ref{Approach}, once our program execution reaches line 1 in the code, a check happens to examine if the allocation is a memory leak.
%which helps us for subsequent tracking.
Given that there is a match in the leak database, we tag the leaked object on line 1, i.e., a special value that corresponds to the leaked object p is added to its shadow memory. This fulfills design component (D3).

\textit{Tracking leaks.}
Once the leaked object is tagged, we want to track its execution path, i.e., we track all code points that the leaked object passes through. 
For example, the code shown in Figure~\ref{Approach} contains calloc function on line 1 for variable p (which we already know is a memory leak). 
Subsequent lines that write to *p (i.e., lines 2 \& 3) are a set of lines that are part of the execution path of the leaked object.
%because they use this allocated memory. 
Hence, we need to collect the stack traces of both lines 2 and 3.
To do so, we utilize the instrumentation (added in step 1) as guard checkpoints surrounding every read and write operations on memory, in order to check whether the read/write happens on a tagged leaked object, by checking the tag value of the object in the shadow memory.
If the shadow memory of the object contains the special value, we record the current program stack in the execution path of the leak.
Since both conditions are satisfied, i.e., line 2 and line 3 are memory writes to variable p and p is already tagged in the shadow memory, we record the stacks related to both lines. Overall, this step implements design component (D4) discussed in previous section. Finally we store the stacks as part of the execution path of p in the leak database. This is the final crucial part of design component (D5).
The output of this step is a list of stacks that represent execution paths of a memory leak, e.g., an execution profile consisting of stacks S1 and S2 (corresponding to lines 2 and 3) is the output after applying this step on our given program, as shown in Figure~\ref{Approach}.
Note that we provide more details about the special value of shadow memory and tracking process in Technical Challenges Section~\ref{sec:implementation}.

\vspace{2mm}
\noindent
\textbf{Step 4: Suggesting fix location for a memory leak.}
In the previous steps, we are able to profile and obtain a set of different execution paths (list of stack traces) along which memory is leaked. 
In this step, we utilize the execution paths to suggest fix locations of a memory leak.
%The goal of this step is to suggest a fix location of the memory leak. 
Ideally, the last code point in the tracked execution path of a memory leak allocation is considered to be a fix location.
For example, in the code shown in Figure~\ref{Approach}, to fix the memory leak on line 1, a developer needs to add a deallocation statement after line 3, as this is the latest line in which the leaked object p is being used.
To identify such a fix location, we identify the last code point on our tracked execution profile. The execution profile for the leak in our example consists of stacks S1 followed by S2. Therefore the last code point in this execution profile is the stack S2. 
% To identify such a fix location, we use the comparison operator $>_s$ (from Section~\ref{sec:design}) for the purpose of finding the last point at which reads and writes operations on leaked object have happened over the test run. 
% We use the operator to compare the obtained stack values.
% In our given example in Figure~\ref{Approach}, applying the operator on the stacks S1 and S2, we find that line 3 follows line 2 in the same function, and hence, we store stack S2 in our leak database as last use point.
\textcolor{black}{Our approach aims to suggest the stack of the last use point as a fix to the leak, and hence, a developer can  benefit from our solution by manually adding a free statement at the suggested point, i.e., the developer needs to deallocate the leak immediately after line 3.}
The solution is further refined over several test runs as more execution paths are uncovered, i.e., we find the last use point over all of the execution traces stored in the leak database.
Note that if the program contains multiple malloc functions, our approach still tackles them because we define a separate execution profile for each individual leaked object.
AddressWatcher also handles scenarios where multiple frees are required for a given memory leak. A detailed explanation of the AddressWatcher algorithm that handles multiple deallocations is described in Technical Challenges Section ~\ref{sec:implementation}.
% A detailed explanation of how we implement the comparison operator is provided in Implementation Section~\ref{sec:implementation}. 
This step enables the realization of the design component (D6).

\section{Technical challenges}
\label{sec:implementation}
In the previous section, we provide a high-level description of how AddressWatcher works to suggest a memory leak fix. 
In this section, we describe specific implementation details of different components of the approach to provide insights into the technicalities of our work. 
We build prototype for AddressWatcher atop ASAN~\cite{b2} and LSAN~\cite{b5} runtime libraries within \verb|gcc| compiler.

%We provide implementation details of our approach under several major headings that we consider when implementing different components of the approach.
%to explain the barriers we encountered while implementing the simple idea in approach overview and how we overcame them.\\
%\mhmd{I think we should not refer too much to the previous steps/section, so that reader do not get confused.}

\vspace{2mm}
\noindent
\textbf{Detecting leaks. } We use a sanitizer for detecting memory leaks - LSAN~\cite{b5}. During compile time, we compile with LSAN option, which add a leak checker module to identify leaks in the program.

% \noindent
% \textbf{Constant memory address offset.} 
% Throughout our approach, we re-execute a given program with multiple testcases. The goal of the first run is to identify memory leaks in the program (step 2, Section~\ref{sec:approachoverview}).
% Subsequent runs aim to track the already identified leaks (step 3, Section~\ref{sec:approachoverview}). In fact, modern systems adopt a technique called ASLR (Address Space Layout Randomisation)~\cite{b46} which randomizes memory addresses on different binary runs to increase the difficulty of attackers for exploiting buffer overflows. 
% Hence, different runs of a binary program would have different (random) memory address offsets. 
% In our approach, it is essential to maintain one constant memory address offset throughout all runs, and therefore, we disable ASLR to maintain consistency of stack traces.\\

\vspace{2mm}
\noindent
\textbf{Implementing leaks database.} A key component in our approach is the leak database (step 2, Section~\ref{sec:approachoverview}).
To implement the leak database, we create a text file to store allocation stacks of the leaks and their execution trace.
%\mhmd{Not sure how important is the naming convention?}
Note that we create a unique leak database for every binary, i.e., the database name is a combination of the instrumented binary name and the instrumented binary directory path.
We also use a special directory to store the leak database. This directory has read and write permissions only for the given user running the binary.
% This ensures that sensitive program data in the leak database is not accessible to other users.

\vspace{2mm}
\noindent
\textbf{Recompilation constraints.} The design of AddressWatcher also takes into account the case when a program binary gets recompiled. 
For example, for the code example shown in Figure~\ref{Approach}, AddressWatcher suggests adding a fix after line 3. Let us assume that a developer agrees with the suggestion and inserts a free statement in line 4.
In this case, if a recompilation happens after, AddressWatcher must stop tracking the memory leak allocation at line 1 because the leak has been fixed.
However, the leak database still has a stack stored in it, suggesting that allocation at line 1 is a leak. 
Hence, recompilation renders all information in the leak database useless. 
Therefore, we label each leak database with the compile time of the corresponding binary, and if a recompilation is detected, AddressWatcher flushes the corresponding binary’s leak database.

\vspace{2mm}
\noindent
\textbf{Implementing leak tagging.} AddressWatcher uses shadow memory to encode information on whether a region of memory is tagged as a leak, i.e., we override allocation functions such as malloc/calloc/realloc, so that AddressWatcher can check if the allocation stack trace of the object belongs to the leak database. 
If so, AddressWatcher tags the allocated object by assigning a special value to its shadow memory before finally allocating the memory.
This special value added to the shadow memory corresponds to setting the higher-order word (16 bits) of the shadow memory to hexadecimal value \textbf{0xe}.
We note that unlike previous approaches~\cite{b53,b2,b41,b54} that utilize shadow memory to detect memory errors, AddressWatcher utilizes shadow memory to tag leaks on different executions and to finally suggest leak fixes. 
% This is a unique design aspect of AddressWatcher.

% We do not modify the lower word which encodes the number of addressable legal bytes in the corresponding application memory, which is required by ASAN.
% Hence, AddressWatcher does not compromise legal bytes encoding performed by vanilla ASAN in order to preserve its buffer overflow detection capability.

\vspace{2mm}
\noindent
\textbf{Implementing leak tracking.} 
% To better understand how AddressWatcher utilizes instrumentation to perform leak tracking, we first explain how ASAN instrumentation works.
AddressWatcher utilizes instrumentation to perform leak tracking. This instrumentation is a modification of ASAN instrumentation~\cite{b2}.
Before reading and writing to allocated variables of the program, ASAN inserts instrumentation. 
The ASAN instrumentation is shown in Listing~\ref{Simplified ASAN instrumentation}. Specifically, in lines 1 and 2, the shadow value of variable \verb|Var| is retrieved by accessing shadow memory. 
% The shadow value encodes information on the bytes that are legally accessible by the running program. 
% ASAN uses a zero value to encode shadow values of application bytes that are legally accessible by the program.
In line 3, the instrumentation checks the shadow value.
If it is not 0, 
% the memory access is illegal (buffer overflow), and 
\verb|Report| function is called on line 4.
We use the function \verb|Report| to track execution paths of leaked objects (step 3, Section~\ref{sec:approachoverview}). 
For tagged leaked objects, the special shadow value is always non-zero (as we modify the upper word of shadow value when tagging leaks). 
Hence, the condition shown on line 3 always succeeds and \verb|Report| is invoked. 
% In AddressWatcher, the original \verb|Report| function is modified to check if the shadow value contains the special value tag.
If the shadow value contains the special value tag, AddressWatcher adds the current program stack to execution path of the tagged leaked object.
% Otherwise (If the special value tag is not observed), AddressWatcher considers it as a buffer overflow just like vanilla ASAN would.
Otherwise (If the special value tag is not observed), AddressWatcher continues execution normally.
% This dual purpose of \verb|ReportAndCrash| to detect both buffer overflows and profile execution paths enables us to use ASAN instrumentation without any modification of its functionality.
% These design choices we have thus adopted ensures that AddressWatcher does not modify the original functionality of ASAN, rather taking their advantages to achieve our goal.

\begin{lstlisting}[language=C++, caption=ASAN instrumentation., label={Simplified ASAN instrumentation}]
    ShadowAddr = ShadowMap(&Var);
    ShadowValue = *ShadowAddr;
    if (ShadowValue != 0)
        Report(&Var);
\end{lstlisting}

\begin{table}[tb!]
  \centering
  \caption{Execution trace of example testcases considered. $A$ is allocation code point for leaked object. $o_1$, $o_2$ and $o_3$ are code points where the leaked object is used.}
  \resizebox{0.6\columnwidth}{!}{%
  \begin{tabular}{l r}
    \toprule
    {Testcase} & {Code points executed} \\
    \midrule
    t1 & $A$, $o_1$\\
    t2 & $A$, $o_1$, $o_2$\\
    t3 & $A$, $o_1$, $o_3$\\
  \bottomrule
\end{tabular}
}
\label{exectrace}
\end{table}

\vspace{2mm}
\noindent
\textbf{Multiple free locations.} There can be multiple locations for free statements required to fix a given  memory leak. AddressWatcher resolves this by analyzing all the tracked execution paths of the memory leak from the leak database.  We illustrate this with an example below. For example, consider a case where an object is allocated at code point $A$ and $o_1$, $o_2$, $o_3$ are different code points where the memory allocation is used. Let us assume there are three testcases t1, t2, and t3 which leaks the same object on different execution paths. We describe the execution trace of each testcase as a list of code points executed by the testcase in Table~\ref{exectrace}. AddressWatcher analyzes each execution trace and filters those execution traces which are a subsequence of another execution trace. A subsequence of an execution trace $E$ is the resulting execution trace when one or more code points are deleted within $E$. For example the execution trace of t1 is a subsequence of the execution trace of t2 (iie. code point $o_2$ must be deleted from execution trace of t2 to obtain execution trace of t1).
We filter execution trace of t1 because if a free is inserted at it's last code point (iie. $o_1$) it will result in a use-after-free vulnerability when testcase t2 is executed. AddressWatcher then obtains the last code point within the  remaining execution traces which are points where the developer is suggested to add the deallocation statements. We note that as the test suite covers more leaked paths, AddressWatcher refines it's fixes by storing relevant execution traces in the leak database for analysis.

\vspace{2mm}
\noindent
\textcolor{black}{\textbf{Custom API deallocations.} AddressWatcher only suggests the location of the last use point of leaked memory. The developer must identify the pointer to the allocated memory from this last use point and manually insert the deallocation statement thereafter. The developer can deallocate this leaked memory at the suggested location in their own desired way using custom APIs. We note that these custom APIs still fundamentally deallocate memory using} \verb|free| \textcolor{black}{ function.}

\section{Evaluation}
\label{sec:evaluation}
In this section, we evaluate our proposed approach.
In particular, we
introduce our evaluation dataset in Section~\ref{sec:data} and
present how fixes suggested by AddressWatcher compare with manual fixes from developers in Section~\ref{sec:res}.
Next, we conduct a comparison with Memfix tool, the state of the art approach for statically fixing memory leaks, in Section~\ref{sec:comparison}. We then present the efficiency evaluation of AddressWatcher in Section~\ref{sec:eval}. \textcolor{black}{Finally, we demonstrate the practicality of AddressWatcher by submitting 25 PRs across 12 diverse open-source software. We discuss these PRs in depth in Section~\ref{sec:OSS}.}

\subsection{Benchmark}
\label{sec:data}
\vspace{-.4mm}
We evaluate AddressWatcher over real world memory leaks that have been fixed by developers in open-source projects.
In particular, we choose to evaluate AddressWatcher over the same benchmark examined by Memfix, a static-based approach for fixing memory leaks in C/C++ programs~\cite{b13}.
The benchmark used by Memfix includes memory leak bugs in five open source packages, namely, \verb|openssh|, \verb|binutils|, \verb|tmux|, \verb|openssl| and \verb|git|~\cite{b32,b31,b34,openssl,git}.

% The selected packages for evaluation represent various domains.
% \verb|openssh| is the primary connectivity tool for remote connectivity through SSH-protocol eliminating eavesdropping and hijacking. 
% \verb|binutils| is a set of programming tools widely used for managing binary programs, object files and assembly source code. 
% \verb|tmux| is an open-source terminal session multiplexer for UNIX-like operating systems. \verb|openssl| provides secure communications over computer networks and \verb|git| is a version control system for software development.
% Table~\ref{Repo stats} presents descriptive statistics on the selected packages.
% Overall, the packages are of interest to the community. 
% From Table \ref{Repo stats}, we can see that each package repository is well-maintained (have a recent date for the latest commit at the paper submission time) and quite large in number of lines of code.

The benchmark in our data comprises 50 fixes of memory leak bugs distributed across the five packages.
% , as shown in Table~\ref{Bug count}. 
There are 10 bug fixes in each of the following repositories - \verb|binutils|, \verb|tmux|, \verb|openssh-portable|, \verb|openssl| and \verb|git|. 
To verify that these bugs are memory leak fixes in the program, we manually analysed them by looking at the corresponding GIT commit, in order to ensure that it discusses fixing a memory leak bug through information in the commit message and the discussion of the corresponding pull-request.

\textcolor{black}{The ground truth for the bug fixes in our benchmark are locations (line numbers) where a developer inserted a} \verb|free| \textcolor{black}{statement for the corresponding memory allocation.
To perform our evaluation, we compare this ground truth with the solution suggested by AddressWatcher.
If AddressWatcher suggests a fix to be added at the same location as the one inserted by the developer (ground truth), we consider that AddressWatcher successfully fixed the memory leak bug. In contrast, if AddressWatcher suggests a fix that has a different location from the ground truth, we consider AddressWatcher to fail fixing the bug.}

%   \begin{table}[tb!]
%   \caption{Statistics on the three packages in our benchmark.}
%   \centering
%  \begin{tabular}{c|c|c}
%     \toprule
%     \thead{Repository name} & \thead{Lines of code} & \thead{Latest commit}\\
%     \midrule
%     binutils & 2,375,596 &17/09/2023\\
%     tmux & 106,528 &17/09/2023 \\
%     openssh-portable & 60,387 & 17/09/2023 \\
%     openssl & 450,982 & 17/09/2023\\
%     git & 255,357 & 17/09/2023\\
%   \bottomrule
% \end{tabular}
% \label{Repo stats}
% \end{table}

\subsection{AddressWatcher Effectiveness}
\label{sec:res}

\iffalse
\begin{figure}[tb!]
     \centering
     \begin{tikzpicture}
\begin{axis}[
    ybar,
    enlargelimits=0.1,  
    ylabel={\# Fixed Memory Leaks},
	xlabel={\ Repository},  
	symbolic x coords={binutils, tmux, openssh},
	ymin=0,
	ytick={0,1,2,3,4,5},
    xtick=data,  
     nodes near coords, % this command is used to mention the y-axis points on the top of the particular bar.  
    nodes near coords align={vertical},  
    legend style={at={(0.5,-0.1)},
	anchor=north,legend columns=-1},
]
\addplot 
	coordinates {(binutils,4) (tmux,3)
		 (openssh,3) };
\addplot 
	coordinates {(binutils,3) (tmux,5) 
		(openssh,5) };
\legend{AddressWatcher,Memfix}
\end{axis}
\end{tikzpicture}
     \caption{Distribution of bugs correctly fixed over different repositories. \mhmd{Not sure if this brings more or less value to the paper. It shows Memfix has better performance. We can show the shared bugs fixed by both and then the independent bugs fixed by each?}}
     \label{Repository Performance}
 \end{figure}
 \fi

%  \begin{table}[tb!]
%   \caption{Number of memory leaks in each package repository.\textcolor{red}{This table can be removed}}
%   \centering
%    \resizebox{0.5\columnwidth}{!}{%
%   \begin{tabular}{c|c}
%     \toprule
%     \thead{Repository name} & \thead{Total bugs}\\
%     \midrule
%     binutils&10\\
%     tmux&10\\
%     openssh-portable&10\\
%     openssl&10\\
%     git&10\\
%      \midrule

%     \textbf{Total}&50\\
%   \bottomrule
% \end{tabular}
% }
% \label{Bug count}
% \end{table}

\begin{table*}[tb!]
 
\end{table*}

Table~\ref{Bug info} presents the distribution of 50 bug fixing commits across the three studied packages in our dataset.
For each fixing commit, Table \ref{Bug info} shows whether AddressWatcher fixes the corresponding bug.
Overall, AddressWatcher fixes a total of 23 bugs (out of 50 bugs) with an accuracy of 46\%.
Breaking down the fixes per package, AddressWatcher fixes 6 bugs in \verb|binutils|, 6 bugs in \verb|tmux|, 3 in \verb|openssh|, 3 in \verb|openssl| and 5 in \verb|git| covering a total of 23 bugs.
That said, AddressWatcher fails to fix the remaining 27 bugs.

%\subsection{Manual Analysis\mhmd{I feel the previous section is too small to be a section on its own. I would suggest merging both 5.2 abd 5.3 as results section.}}
We manually look at each one of the 27 bugs to understand the context of the affected code and the reason why AddressWatcher could not suggest the same fix as the one developers add in the code.
Table \ref{Bug info} summarizes the reasons as to why AddressWatcher failed to fix a given memory leak bug.
Below, we provide more details about each reason.\\

\begin{table}[tb!]
  \caption{List of 50 bug fixing commits in our benchmark, per package. For each commit, we present: 1) whether AddressWatcher can fix the bug, 2) the reason AddressWatcher fails to fix the bug, 3) whether Memfix can fix the bug.
  }

\resizebox{0.49\textwidth}{!}{
  \begin{tabular}{c|c|c|c|c}
    \toprule
    \thead{Repository} & \thead{Leak\\No.} &  \thead{Fixed \\by AW\\ ($\checkmark$/$\times$)} & \thead{AW Failure\\reason} & \thead{Fixed \\by MF?\\  ($\checkmark$/$\times$)}\\
    \midrule
    \multirow{10}*{binutils}  & \href{https://github.com/bminor/binutils-gdb/commit/be74fad95edc8827516e144cf38d135b503249cd}{1} & $\checkmark$ & - & $\checkmark$\\
                             & \href{https://github.com/bminor/binutils-gdb/commit/3cfd3dd0956fe854a07795de12c1302ecabbd819}{2} & $\checkmark$ & - & $\checkmark$\\
                             &
                             \href{https://github.com/bminor/binutils-gdb/commit/3f2a3564b1c3872e4a380f2484d40ce2495a4835}{3} & $\checkmark$ & - & $\checkmark$\\
                             &  \href{https://github.com/bminor/binutils-gdb/commit/aba19b625f34fb3d61263fe8044cf0c6d8804570}{6} & $\checkmark$ & - & $\checkmark$\\ 
                             & \label{examplecommit} \href{https://github.com/bminor/binutils-gdb/commit/b55ec8b676ed05d93ee49d6c79ae0403616c4fb0}{4} & $\checkmark$ & - & $\times$\\
                             &\label{examplecommit} \href{https://github.com/bminor/binutils-gdb/commit/848ac659685fba46ce8816400db705f60c8040f7}{5} & $\checkmark$ & - & $\times$\\
                             &
                             \href{https://github.com/bminor/binutils-gdb/commit/a26a013f22a19e2c16729e64f40ef8a7dfcc086e}{7} & $\times$ & Error path & $\times$\\
                             &
                             \href{https://github.com/bminor/binutils-gdb/commit/f978cb06dbfbd93dbd52bd39d992f8644b0c639e}{8} & $\times$ & Code Organization & $\times$\\
                             & \href{https://github.com/bminor/binutils-gdb/commit/7ed1acafa0b5d135342f9dcc0eb0387dff95005a}{9} & $\times$ & Code Organization & $\times$\\
                              &  \href{https://github.com/bminor/binutils-gdb/commit/8a286b63457628b0a55d395f14005f254512e27d}{10} & $\times$ & Error Path & $\times$\\\midrule 
                            
    \multirow{9}*{tmux}      & \href{https://github.com/tmux/tmux/commit/c363c236aaea5b7a879493d8f3c85bead546f063}{1} & $\checkmark$ & - & $\checkmark$\\
                             & \href{https://github.com/tmux/tmux/commit/1e0eb914d945e0f287716d56669d0de409e86e59}{2} & $\checkmark$ & - & $\checkmark$\\
                             &
                             \href{https://github.com/tmux/tmux/commit/d566c780e54010112d499707cd80a594144d1a89}{3} & $\checkmark$ & - & $\checkmark$\\
                             & \href{https://github.com/tmux/tmux/commit/2c9bdd9e326723fb392aed4d8df12cba7ef34f1f}{4} & $\checkmark$ & - & $\checkmark$\\  
                             & \href{https://github.com/tmux/tmux/commit/51ac2a3202d55c439976ecce49229e35865c7ebd}{5} & $\checkmark$ & - & $\times$\\
                             & \href{https://github.com/tmux/tmux/commit/6daf06b1ad61f67e9f7780d787451b9b5f82dd43}{6} & $\checkmark$ & - & $\times$\\
                             &  \href{https://github.com/tmux/tmux/commit/933929cd622478bb43afe590670613da2e9ff359}{7} & $\times$ & Error Path & $\checkmark$\\
                             & \href{https://github.com/tmux/tmux/commit/7340d5adfdc8cc6d845a373f3e0d59bfd10a45d1}{8} & $\times$ & Error Path & $\times$\\
                             &  \href{https://github.com/tmux/tmux/commit/189017c078b7870c18ced485c1fd99f65fcc4801}{9} & $\times$ & Error Path & $\times$\\
                             & \href{https://github.com/tmux/tmux/commit/5acee1c04ed38afd6a32da4a66e6855ccdc52af3}{10} & $\times$ & Error Path & $\times$\\
                             \midrule 
                            
    \multirow{10}*{openssh-portable}  
                             & \href{https://github.com/openssh/openssh-portable/commit/0d6771b4648889ae5bc4235f9e3fc6cd82b710bd}{1} & $\checkmark$ & - & $\checkmark$\\
                             &
                             \href{https://github.com/openssh/openssh-portable/commit/aae07e2e2000dd318418fd7fd4597760904cae32}{2} & $\checkmark$ & - &  $\times$\\
                             &  \href{https://github.com/openssh/openssh-portable/commit/393920745fd328d3fe07f739a3cf7e1e6db45b60}{3} & $\checkmark$ & - & $\times$\\
                             & 
                            \href{https://github.com/openssh/openssh-portable/commit/b2afdaf1b52231aa23d2153f4a8c5a60a694dda}{4} & $\times$ & Error Path & $\checkmark$ \\
                             & 
                             \href{https://github.com/openssh/openssh-portable/commit/64a89ec07660abba4d0da7c0095b7371c98bab62}{5} & $\times$ & Error Path & $\checkmark$\\
                             & \href{https://github.com/openssh/openssh-portable/commit/165bc8786299e261706ed60342985f9de93a7461}{6} & $\times$ & Error Path & $\checkmark$\\
                             & \href{https://github.com/openssh/openssh-portable/commit/66d2e229baa9fe57b86}{7} & $\times$ & Error Path & $\checkmark$\\
                             &
                             \href{https://github.com/openssh/openssh-portable/commit/4f7cc2f8cc861a21e6dbd7f6c25652afb38b9b96}{8} & $\times$ & Weak test suite & $\checkmark$\\
                             & \href{https://github.com/openssh/openssh-portable/commit/a63cfa26864b93ab6afefad0b630e5358ed8edfa}{9} & $\times$ & Error Path & $\times$\\
                             &
                             \href{https://github.com/openssh/openssh-portable/commit/e52a260f16888ca75390f97de4606943e61785e8}{10} & $\times$ & Weak test suite &  $\times$\\
\midrule                             
\multirow{10}*{openssl}
                            & \href{https://github.com/openssl/openssl/commit/8abeefeccc4cfbfba9b5ebfc7604fe257a97317a}{1} & $\checkmark$ & - &  $\checkmark$\\
                            & \href{https://github.com/openssl/openssl/commit/af6de400b49c011600cfd557319d1142da6e5cbd}{2} & $\checkmark$ & - &  $\times$\\
                            & \href{https://github.com/openssl/openssl/commit/04761b557a53f026630dd5916b2b8522d94579db}{3} & $\checkmark$ & - &  $\times$\\
                            & \href{https://github.com/openssl/openssl/commit/85155346b3ca2dcdecf018dc8db9df94ceebeb0d}{4} & $\times$ & Error Path &  $\checkmark$\\
                            & \href{https://github.com/openssl/openssl/commit/44f19af7434cdb996f1ce11789150baa07db27e6}{5} & $\times$ & Error Path & $\checkmark$\\
                            & \href{https://github.com/openssl/openssl/commit/59099d6b8a3aec77f7d9f310ebf8e31b09c2d518}{6} & $\times$ & Error Path &  $\checkmark$\\
                            & \href{https://github.com/openssl/openssl/commit/62b0a0dea612e3683c6bd4bef359fceda00238e8}{7} & $\times$ & Error Path &  $\checkmark$\\
                            & \href{https://github.com/openssl/openssl/commit/b6306d8049b04dca7fa738a86c892c43ba6a5fc4}{8} & $\times$ & Error Path &  $\times$\\
                            & \href{https://github.com/openssl/openssl/commit/918a27facd3558444c69b1edbedb49478e82dff5}{9} & $\times$ & Error Path &  $\times$\\
                            & \href{https://github.com/openssl/openssl/commit/9561e2a169f499f8346ffdd7541bc4e3d81d6711}{10} & $\times$ & Weak test suite &  $\times$\\
\midrule
\multirow{10}*{git}
                            & \href{https://github.com/git/git/commit/cb7b29eb67772d08e2365ed07ede9d954d0344c1}{1} & $$\checkmark$$ & - &  $$\checkmark$$\\
                            & \href{https://github.com/git/git/commit/dd1055ed594f8fef18779cce3cd921c4ac66cf9c}{2} & $$\checkmark$$ & - &  $\times$\\
                            & \href{https://github.com/git/git/commit/a452d0f4bae99c9acef6f7db75f6f1d922618732}{3} & $$\checkmark$$ & - &  $\times$\\
                            & \href{https://github.com/git/git/commit/e336bdc5b9bcb62982da9708dfb6e68150de72a3}{4} & $$\checkmark$$ & - &  $\times$\\
                            & \href{https://github.com/git/git/commit/fc5c40bb2bb1921f3bdfa55c1d846dc080c356b2}{5} & $$\checkmark$$ & - &  $\times$\\
                            & \href{https://github.com/git/git/commit/f4e45cb3eb6fad4570ff63eecb37bae8102992fc}{6} & $\times$ & Weak test suite &  $\times$\\
                            & \href{https://github.com/git/git/commit/dcb572ab94f83a1a857d276fcebff5700077f2b7}{7} & $\times$ & Weak test suite &  $\times$\\
                            & \href{https://github.com/git/git/commit/4da72644b768b0491110a8ba0aa84d32b6bde41c}{8} & $\times$ & Error Path &  $\times$\\
                            & \href{https://github.com/git/git/commit/851e1fbd01250f56a6e479e1addada220a56e1f7}{9} & $\times$ & Error Path &  $\times$\\
                            & \href{https://github.com/git/git/commit/afbb8838b7d4d1887da1e1871f8e9ccd01ae1138}{10} & $\times$ & Error Path &  $\times$\\     
\bottomrule
\end{tabular}
}
\label{Bug info}
\end{table}

\noindent
\textbf{Error paths (20 cases).} \label{Errorpathexplanation}
The most common reason of AddressWatcher failing to provide a correct fix is when the program contains error paths. An error path refers to a program path where an abrupt return happens under abnormal situations (incorrect arguments, low memory, etc.). 
Listing~\ref{Fail Result} shows a real example of such a case for a bug in \verb|binutils|~\cite{b38}. In the Listing, the error path could be triggered in case the user does not supply the correct arguments to the program.
The allocation on line 1 is leaked only on the error path (when \verb|goto FAIL| is executed on line 2) and not on the non-error path (for the non-error path, there is already a deallocation on line 4). 
Hence, a developer would insert the fix in the error path before the \textit{exit} keyword on line 8. 
In such cases, AddressWatcher fails to provide the same fix as the developer's solution. 
This is because there is no read/write operation to the leaked object in the error handling routine, and hence, our approach is not able to track the leaked object. 
In such cases, AddressWatcher suggests a fix immediately after the allocation on line 1.\\
%To correctly consider these cases, AddressWatcher needs to perform some form of static analysis to fix error paths as well.

%breaklines=true

\begin{lstlisting}[language=C++, caption=Code snippet from binutils showing AddressWatcher failure due to an error path~\cite{b38}., label={Fail Result}]
    char *p = malloc(10); // AddressWatcher suggests fix after allocation
    if(argv == NULL)  goto FAIL; // Memory leak
        *p = 'a' // Non-error path
    free(p);
    return 0;
    FAIL:
    // Error path
    // Developer inserts fix here
    exit(1);
\end{lstlisting}

\noindent
 \textbf{Weak test suite (5 cases).} Other cases where AddressWatcher could not suggest an accurate fix is when the program test coverage is weak or insufficient. This is because AddressWatcher relies upon the test suite to track leaks on new paths.
 % Listing~\ref{TestC} discusses an example based on the error path code.
 % In the example, the leaked variable p is used on the error handling routine on line 6. The developer would insert the free after this use on line 7. AddressWatcher is now capable of fixing the bug because the leaked variable is used in the error path. 
 % However, in this specific code, no testcase executes the error path, i.e., no test case considers providing invalid arguments, and hence, AddressWatcher was not able to fix the bug. 
 % AddressWatcher would suggest the fix immediately after the allocation in this case.
 That said, with a better testsuite coverage, AddressWatcher is still able to suggest the correct fix for these cases.
 \\

\begin{lstlisting}[float, language=C++,caption=Code snippet from binutils showing AddressWatcher failure due to code organization~\cite{b37}., label={codeorganization}]
    struct btrace {
        char* p;
        char* q;
    }
    void btrace_alloc (struct* X) {
        X = malloc(sizeof(X));
        X->p = malloc(10);
        X->q = malloc(10);
    }
    void btrace_clear (struct* X) {
        // Developer adds free(X->p) here
        free(X->q); 
        free(X);
    }
    int main () {
        struct btrace* X;
        btrace_alloc(X); // Allocate X,p,q
        doSomething1(X->p); // Use X->p
        // AddressWatcher suggests fix for X->p here
        doSomething2(X->q); // Use X->q
        btrace_clear(X); // Deallocate X,q
        return 0;
    }
\end{lstlisting}

\noindent
 \textbf{Code organization (2 cases).} Code organization refers to developer's decisions in structuring the code in a certain way to promote general code reliability and adaptability to future changes. This can lead developers to provide fixes that are not always immediately after the last use of the allocated object.
 Such a case is seen in a bug that affects the package \verb|binutils| ~\cite{b37}, which is illustrated in Listing~\ref{codeorganization}. 
 As the Listing shows, there is a memory leak originating from allocation of variable p within struct X of type \verb|btrace|. 
 The developer constructs two separate routines for allocation and deallocation of the struct X and its inner variables, one is called \verb|btrace_alloc| (on line 5) which is used for allocation, and the other is called \verb|btrace_clear| (on line 10) for deallocation.  
 X\verb|->|p is allocated in \verb|btrace_alloc| and its last use occur in routine \verb|doSomething1|. Hence, AddressWatcher suggests the last use point after routine \verb|doSomething1| as the fix.  However, the developer inserts the fix on line 11 in \verb|btrace_clear| before the struct X itself is freed. This could have been done for general code reusability reasons for future commits. We count this as a negative case for AddressWatcher because we are comparing with the ground truth being the location where the developer inserts the fix. 
 However, we should note that the solution suggested by AddressWatcher is more optimal (freeing memory earlier) than the developer's fix.

\subsection{Dynamic (AddressWatcher) Fixing Compared to Static (Memfix) Fixing}
\label{sec:comparison}
In this section, we compare AddressWatcher to Memfix, the \textit{state-of-the-art} approach for statically fixing memory leaks~{\cite{b13}}. %Memfix is an open source tool for fixing memory leaks through static analysis. 
The goal of our comparison with a static-based analysis approach is to expose the gap between static and dynamic analysis for fixing memory leak bugs.

Memfix attempts to fix memory leaks by identifying a set of free statements that deallocates all objects without causing double-free or use-after-free. 
It utilizes the insight that identifying set of deallocation statements corresponds to an exact cover problem on a variant of typestate static analysis.
It uses a SAT solver to solve the exact cover problem~\cite{b10}. 
Then, all frees that are present in these paths are removed, and new free statements are added.

\begin{table}[tb!]
  \centering
  \caption{Memory leak fixes by AddressWatcher \& Memfix, per package.}
  \resizebox{\columnwidth}{!}{%
  \begin{tabular}{l r r r}
    \toprule
    \thead{Repository name} & \thead{Total bugs} & \thead{\# Fixes by AddressWatcher} & \thead{\# Fixes by Memfix} \\
    \midrule
    binutils&10&6&4\\
    tmux&10&6&5\\
    openssh-portable&10&3&6\\
    openssl&10&3&5\\
    git&10&5&1\\
    \midrule
    Total&50&23&21\\
  \bottomrule
\end{tabular}
}
\label{Bug count fix}
\end{table}

We compare AddressWatcher to Memfix over the same packages examined in the benchmark.
\textcolor{black}{We set up a virtual machine provided by Memfix and run Memfix with its default configuration over the bugs in the benchmark~\cite{b47}. We run Memfix without any time constraints and compare results with AddressWatcher.}
% We used the virtual machine shared by Memfix authors where the tool was already setup and ran the tool over the bugs in the benchmark.
%
% After running the virtual machine on the dataset, we find that AddressWhatcher could fix x bugs while Memefix could provide proper fixes for 12 bugs.
% \mhmd{A good way to start presenting the results is to use a figure or table. Then, describe using that.}

Table ~\ref{Bug count fix} shows the number of memory leak fixes by both Memfix and AddressWatcher, per package repository. 
From the Table, we can see that, of the 50 bugs examined in our benchmark, AddressWatcher fixes 23 bugs while Memfix fixes 21 bugs.

% AddressWatcher fixes 10 bugs, Memfix fixes 12 bugs, and 10 bugs are fixed by neither AddressWatcher nor Memfix.

\begin{figure}[tb!]
    \centering
    \resizebox{0.9\columnwidth}{!}{
    \begin{tikzpicture}  
  
\begin{axis}  
[  
    ybar,  
    enlargelimits=0.15,  
    ylabel={\# Memory leak fixes}, % the ylabel must precede a # symbol.  
    xlabel={\ Tool},  
    symbolic x coords={AW only, MF only, AW \& MF, No tools}, % these are the specification of coordinates on the x-axis.  
    xtick=data,  
     nodes near coords, % this command is used to mention the y-axis points on the top of the particular bar.  
    nodes near coords align={vertical}, 
    % xmin=0,
    % ymin=0
    ]  
\addplot coordinates {(AW only,12) (MF only,10) (AW \& MF,11) (No tools,17) };  
  
\end{axis}  
\end{tikzpicture}  
}

    \caption{Distribution of memory leak fixes by AddressWatcher (AW) and Memfix (MF).}
    \label{Result}
\end{figure}
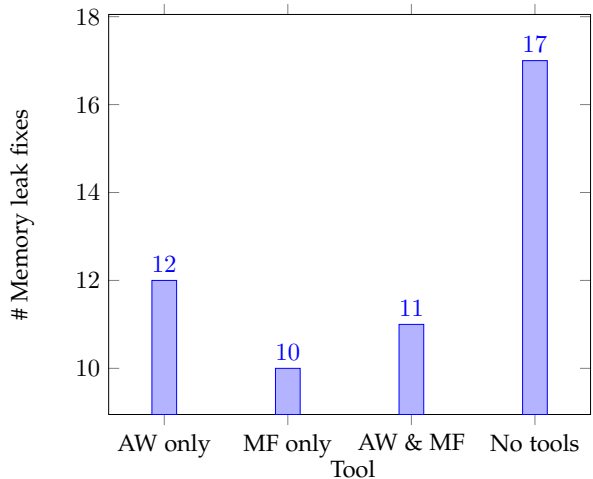

% We have already discussed failure reasons for AddressWatcher in Section~\ref{sec:res}.
% We now provide further investigation that sheds light on when Memfix succeeds or fails to fix the leak.

\textcolor{black}{We now aim to understand the intersection of memory leak fixes introduced by AddressWatcher and Memfix, i.e., bugs independently and jointly fixed by both tools. We outline the distribution of memory leak fixes in Figure~\ref{Result}. The figure shows that AddressWatcher fixes 12 bugs independently of Memfix, i.e., Memfix fails to suggest fixes for these bugs. Also, we can see that there are 10 bugs fixed by Memfix independently of AddressWatcher. Both AddressWatcher and Memfix jointly provide proper fixes for 11 bugs. Still, 17 bugs in the examined repositories were not fixed by both AddressWatcher and Memfix. Next, we provide further investigation into those bugs which are not fixed by atleast one approach (static or dynamic). We describe these in the following three scenarios (S1 – S3):}\\

% \noindent
% \textbf{S1. Memory leak bugs fixed by both AddressWatcher and Memfix.} \textcolor{black}{For the 11 cases solved by both AddressWatcher and Memfix, we find that the allocated and leaked variable is largely limited in scope within a procedure and has limited leaked paths. All of these leaked paths are also easily covered by simple test cases. Hence, both approaches are able to fix them.}\\

\noindent
\textbf{S1. Memory leak bugs fixed by AddressWatcher only.} \textcolor{black}{As shown previously, AddressWatcher provides 12 fixes independently of Memfix. Among them, Memfix fails to generate and solve constraints for its SAT solver in 6 cases due to the complex nature of these interprocedural leaks. Memfix also fails in 3 cases where leaked memory is reallocated (with realloc). Finally, Memfix fails in the remaining 2 cases where recursion is not supported. Additionally, in one case, the Memfix application crashes without providing a reason for the crash. In contrast, AddressWatcher succeeds in such cases because it is a dynamic approach where it relies on a test suite to uncover all leaked paths.}\\

\noindent
\textbf{S2. Memory leak bugs fixed by Memfix only.} \textcolor{black}{To understand the nature of the 10 bugs fixed by Memfix independently of AddressWatcher, we manually look at these cases, trying to understand the nature of each case. We find that most of these bugs are related to error path cases (9 cases). Another failure case is due to a weak test suite not covering the leaked path. We previously describe the context of all such cases where AddressWatcher fails in Section}~\ref{sec:res}.\\

\noindent
\textbf{S3. Memory leak bugs not fixed by both AddressWatcher and Memfix.} \textcolor{black}{For the 17 cases that both approaches fail to provide a fix, our manual investigation shows that the majority of the bugs that AddressWatcher could not fix are due to error paths (11 cases). There are 4 memory leaks where AddressWatcher fails to fix due to a weak test suite, and the remaining 2 cases are due to code organization issues. Memfix encounters issues in specific scenarios: it fails when dealing with reallocated memory (10 cases), cannot generate constraints for its SAT solver in certain complex interprocedural leaks (6 cases), or when dealing with function pointers (1 case).}\\

% Finally, the last case is due to the inaccuracy of the comparison operator employed by AddressWatcher.
% We previously describe the context of all such cases where AddressWatcher fails in Section~\ref{sec:res}.\\
%Memfix is able to fix these bugs because of smaller interprocedural paths considered and the fact that it can track the leak down an error path.\\
%as shown in Listing \ref{Fail Stack comparison}.

% \noindent
% \textbf{Memory leak bugs not fixed by both AddressWatcher and Memfix.}
% For the 17 cases that both approaches fail to provide a fix, our manual investigation shows that the majority of the bugs that AddressWatcher could not fix are due to no use of leaked variables on error paths (11 bugs).
% There are 5 memory leaks where AddressWatcher fails to fix due to a weak test suite, and the remaining 2 cases are due to code organization issues. 
% In some of the cases for Memfix, again the SAT solver is unable to find a solution when there are several leaked program paths.
% Memfix was also found not supporting recursion, and the \verb|realloc| method of allocation in their static analysis, which is not the case in dynamic approaches like AddressWatcher.\\

\noindent
\textcolor{black}{To sum up, our comparison demonstrates that AddressWatcher complements Memfix by tackling fixes along non-error paths, where Memfix errors out. Moreover, note that Memfix does not support recursion, function pointers and memory reallocation in their static analysis, which is not the case in dynamic approaches like AddressWatcher. These tools are, hence, complementary in nature and can be used together to fix 33 real-world leaks in our benchmark.}

\subsection{AddressWatcher Efficiency}
\label{sec:eval}
 \textcolor{black}{In this section, we evaluate the efficiency of AddressWatcher. AddressWatcher utilizes LSAN to detect leaks. Then, in subsequent testcases, it profiles those leaks and suggests final fix locations. Therefore, for Addresswatcher to provide the best solution, each testcase must be run twice. This is the runtime with AddressWatcher instrumentation.  In order to calculate AddressWatcher overhead, we must first establish the time taken to run those testcases without any profiling instrumentation or LSAN. Hence, our baseline is the runtime of the binary with testcases run twice without any instrumentation. Then, the overhead is calculated as follows:
 $((T_I  - T_{WI}) / T_{WI}) * 100$ where $T_I$ is the time taken to execute testcases twice with instrumentation and $T_{WI}$ is the time taken to execute testcases twice without instrumentation.
 The result of this analysis is shown in Table~\ref{AddressWatcher efficiency}.}

 \textcolor{black}{Table~\ref{AddressWatcher efficiency} indicates the average runtime required by AddressWatcher to suggest fixes in a repository. The table also shows the maximum runtime to suggest the leak fix in that repository. Overall AddressWatcher has a performance overhead of 142-178\% over the baseline. This clearly shows the efficiency of AddressWatcher in suggesting memory leak fixes. Additionally, AddressWatcher is a debugging tool, not a runtime checker, and is not intended to be deployed on production servers. Therefore, it is most useful when the code is in the testing phase.}

% ASAN performance overhead can be a maximum of 2x base program~\cite{b2}, while LSAN adds almost no performance overhead~\cite{b44}. AddressWatcher modifies ASAN operations at runtime : (1) checking allocation in leaks database before tagging, and (2) tracking leaks using modified ASAN instrumentation.

\begin{table}[tb!]
  \centering
  \caption{Efficiency of AddressWatcher (AW).}
  \resizebox{0.45\textwidth}{!}{%
  \begin{tabular}{l r r r}
    \toprule
    \thead{Repository} & \thead{\% AW \\ overhead} & \thead{AW avg \\run \\time (s)} & \thead{AW max \\run \\time (s)}\\
    \midrule
    binutils & 178.24 & 0.53 & 0.62\\
    tmux & 160.69 & 0.51 & 0.54\\
    openssh-portable & 142.75 & 0.52 & 0.75\\
    openssl & 147.32 & 0.50 & 0.58\\
    git & 171.36 & 0.48 & 0.60\\
  \bottomrule
\end{tabular}
}
\label{AddressWatcher efficiency}
\end{table}

\subsection{Practicality of AddressWatcher in Open Source}
\label{sec:OSS}

\textcolor{black}{In this section, we demonstrate the practicality and relevance of AddressWatcher to the open-source community. We identified 12 prominent C repositories with memory leak problems using the Github search functionality. We then used the Leaksanitizer tool to detect various memory leaks.  We then submitted 25 Pull Requests to these prominent open-source projects that are affected by memory leak bugs.  
We submitted these PRs by inserting deallocation statements at the locations specified by the AddressWatcher suggestions.} 

\textcolor{black}{Overall, among these 25 PRs, 21 were merged and 4 are pending approval. The merged PRs led to 5 open github issues on memory leaks being resolved~\cite{iniparser-issue,h2o-issue,snmp-issue,whois-issue,nanonng-issue}. One of our merged patches was so critical that it resulted in a new version of the calc software (calc v2.15.0.6) being released~\cite{calc-Release}. Additionally, 3 of our PRs even triggered lively discussion on the coturn repository on how to improve memory safety in the future by upgrading to C++17~\cite{coturn-disc}. 
Below, we provide insights into the context of a subset of these PRs, highlighting their importance and the impact they have on the analyzed projects:}

\begin{lstlisting}[language=C++, caption=whois PR showing AddressWatcher leak fix~\cite{whois1}., label={whois-listing}]
    sockfd = openconn(server, port);
    free(server); // AddressWatcher fix
    server = do_query(sockfd, query_string);
\end{lstlisting}

\begin{lstlisting}[language=C++, caption=h2o PR showing AddressWatcher leak fix~\cite{h2o1}., label={h2o-listing}]
    if (getsockopt(fd, IPPROTO_TCP, TCP_CONGESTION, buf, &buflen) == 0) {
        return h2o_iovec_init(buf, strlen(buf));
    }
    free(buf); // AddressWatcher fix
\end{lstlisting}

\begin{lstlisting}[language=C++, caption=neomutt PR showing AddressWatcher fix~\cite{neomutt1}., label={neomutt-listing}]
    // Displayed if appending to trash fails when syncing or closing a mailbox
    if (mutt_append_message(m_trash, m, e, NULL, MUTT_CM_NO_FLAGS, CH_NO_FLAGS) == -1)
    {
        mx_mbox_close(m_trash);
        mailbox_free(&m_trash); // AW fix
        return -1;
    }
\end{lstlisting}

\begin{enumerate}
    \item \textcolor{black}{Calc (1 PR  merged): Calc~\cite{calc} is an interactive calculator that can be easily programmed for difficult or long calculations. It has been used before to calculate the largest known non-Mersenne prime: $391,581*2^{216,193}-1$. Calc is now being used to calculate other large primes over years of execution. We merged a PR in this repository~\cite{calc-pr} and were given credit for fixing a “long-standing memory leak” on their discussion forum ~\cite{calc-Release} and CHANGES document~\cite{calc-readme}. The repository owners noted that the merged fix addresses a memory leak in calc that previously caused crashes due to gradual memory leak buildup over years of execution, hindering the discovery of larger primes. Additionally, a new version of calc was immediately rolled out (calc v2.15.0.6) after merging the PR indicating the high impact of the fix~\cite{calc-Release}.}
    \item \textcolor{black}{Radare2 (10 PRs merged): Radare is a library used to ease binary reverse engineering tasks and binary exploitation~\cite{radare}. We submitted 10 PRs fixing leaks detected while analyzing ``ls" binary and by running the visual mode of radare~\cite{radare1,radare2,radare3,radare4,radare5,radare6,radare7,radare8,radare9,radare10}. Three of these PRs fixed leaks across multiple locations as well~\cite{radare5, radare3, radare4}. All 10 were merged. In one PR, the author was even impressed with a fix saying ``whoa nice spot!"~\cite{radare10}.}
    \item \textcolor{black}{CoTurn (3 PRs merged): Coturn is an implementation of a VoIP media traffic NAT traversal server and gateway~\cite{coturn}. We submitted 3 PRs to this repository and all 3 were merged~\cite{coturn1,coturn2,coturn3}. The PRs that we submitted initiated a long discussion on how to migrate project code to C++17 to improve memory safety using destructors~\cite{coturn-disc}. Later after submitting the PRs, the first author was approached by mail for a consultancy position.}
    \item \textcolor{black}{Net-snmp (2 PRs merged): Net-snmp is a widely used protocol for monitoring the health of network equipment~\cite{snmp}. We fixed 2 leaks identified by fuzzers when the '-a' argument of net-snmp is used incorrectly~\cite{snmp1,snmp2}. Both PRs were merged resolving an open issue~\cite{snmp-issue}.}
    \item \textcolor{black}{Neomutt (1 PR merged): NeoMutt is a command line mail reader~\cite{neomutt}. We submitted a PR fixing a memory leak when certain mail is added to trash~\cite{neomutt1}, as shown in Listing~\ref{neomutt-listing}. In this case, the trash mailbox is closed, but the memory associated with it is not freed. AddressWatcher is able to locate the line that closes the mailbox as the last use location and suggests a free after this line. The PR fixing the leak was merged and the authors were congratulated for a "good spot".}
    \item \textcolor{black}{Whois (1 PR merged): Whois is a UNIX command-line utility used to make WHOIS protocol queries~\cite{whois}. In the repository, a pointer to allocated memory} \verb|server| \textcolor{black}{is reassigned without being freed, as shown in Listing ~\ref{whois-listing}. AddressWatcher is able to detect this last line where the leaked memory is used and suggests the free after this line. We submitted a PR fixing the leak that is merged~\cite{whois1} and resolved another issue~\cite{whois-issue}.}
    \item \textcolor{black}{h2o (1 PR merged): h2o is a really fast HTTP server~\cite{h2o}. In Listing ~\ref{h2o-listing}, a leak occurs when allocated memory in} \verb|buf| \textcolor{black}{is used to change socket options through} \verb|getsockopt|. \textcolor{black}{If setting socket options fails, the allocated memory is never used and never freed. AddressWatcher identifies last use location to be the} \verb|getsockopt| \textcolor{black}{call.  We submitted a PR that frees the memory after the branch condition, which has been merged~\cite{h2o1} and resolved another issue~\cite{h2o-issue}.}
    \item \textcolor{black}{Iniparser (1 PR merged): This library offers parsing of ini files from C~\cite{iniparser}. We submitted a  PR~\cite{iniparser1} which project owners mentioned to be more comprehensive than another PR submitted by a different user~\cite{iniparser-otheruser}. Our PR was merged in the repository leading to an open issue being resolved~\cite{iniparser-issue}.}
    \item \textcolor{black}{NanoNNG (1 PR merged): NanoNNG solves common recurring messaging problems for IoT devices, such as RPC-style request/reply~\cite{nanonng}. We merged a PR that fixed a memory leak that occurred on reading configuration files with repeating JSON fields~\cite{nanonng1}. This also led to an open issue being closed~\cite{nanonng-issue}.}
\end{enumerate}

We also submitted 4 PRs in 3 other repositories. We submitted 2 PRs in yasm repository~\cite{yasm1,yasm2}, 1 PR in shc repository~\cite{shc1}, and 1 PR in hackem repository~\cite{hackem1}. However, these PRs are still under review and are yet to be merged. 
% Overall, the strength of AddressWatcher lies in its automatic capability to localize leak-fix locations across 25 PRs.

\section{Discussion}
% In the previous section~\ref{sec:evaluation}, we discuss cases where AddressWatcher fails to fix memory leaks. 
This section discusses how certain aspects impact the performance of AddressWatcher when suggesting fix locations.

\vspace{2mm}
\noindent
\textbf{Code coverage.}
\textcolor{black}{AddressWatcher fundamentally relies on a test suite to provide a fix for memory leaks.
Hence, if the test suite does not cover all paths of a leaked object, the suggested solution might not be optimal.
For example, a leaked object may be used through a path that is not covered by the test suite and such a path could be located in a point that is deeper in the program flow than the point we suggest for the fix.
Consequently, the suggested fix may lead to a use-after-free if the test coverage is poor. We note that this is a limitation of all existing dynamic analysis techniques~\cite{b52,b45,b12}.
% Still, AddressWatcher provides a highly relevant analysis for developers to help them find locations where the leaked object is last used, saving significant amount of developer time.
Possible future work to mitigate this in AddressWatcher would be to fuzz the program to obtain a high-quality test suite. This is an orthogonal problem that will further improve the results of AddressWatcher. AddressWatcher is the framework that utilizes these test cases to obtain memory leak fixes.
}

\vspace{2mm}
\noindent
\textbf{Multi-threading support.} 
In the case of a multi-thread program, each thread could lead down a different execution path for a given leak. 
Consider the case where a memory leak has two different execution paths. Thread 1 explores the first execution path and thread 2 explores the second execution path. That is, each thread can lead to a different solution for the same leak.
To mitigate this, we implement a concurrency lock to prevent threads from writing to the database simultaneously.
In the case that thread 1 completes before thread 2, thread 1 first writes its solution to the leak database. 
When thread 2 is about to die, it reads the previous solution in the leak database and compares it with its own solution using the comparison operator. 
Based on the comparison outcome, thread 2 will store the best solution, i.e., the solution that represents a later point in the program path will be stored in the leak database.

\section{Related Work}
Two lines of work are closely related to AddressWatcher:
(1) techniques for memory leak detection, and
(2) techniques for memory leak fixing.
% and (3) studies on garbage collection.
In the following, we discuss the related work and reflect on how the work compares with our work.
%We first describe techniques for memory leak detection. Then we discuss techniques for memory leak fixing including garbage collectors.

\subsection{Approaches for Memory Leak Detection}

\noindent
\textbf{Static approaches.}
A plethora of recent work focused on
detecting memory leaks statically~\cite{b4,b1,b3, Li2020Nov, Cao2022Mar, Fan2019May}.
% For example, Cherem et al. and Sui et al. propose Fastcheck~\cite{b4} and Saber~\cite{b1}, respectively.
% These techniques perform pointer value-flow analysis
% from allocations to deallocation sites.
% More specifically,
% they use a sparse graph representation
% to capture pointer define-use chains.
% Memory leaks detection is hence reduced to
% solving graph reachability queries on the constructed graph.
\textcolor{black}{Static analysis approaches such as Smoke~\cite{Fan2019May} suffers from imprecision in detecting bugs. This can be due to approximations in underlying pointer analysis, lack of library specifications, infeasible paths due to complex arithmetic in branch conditions, and other cases such as recursion, function pointers, etc. Therefore, the common problem of most static approaches is that they incur a high rate of false positives in detecting bugs, and this translates to limitations in bug fixing.}
% Sparrow presented by Jung et al.~\cite{b3} is another memory leak detection approach that investigates common memory leak-related behaviors in real-world C programs.
% They categorise memory leak-related effects of procedures into eight broad classes.
% These classes are instantiated while analyzing call sites of the procedure to identify potential memory leaks.

\vspace{2mm}
\noindent
\textbf{Dynamic approaches.} 
Dynamic approaches for memory leak detection have been discussed broadly. 
For example, 
LeakSanitizer (LSAN) is a tool by Google that performs dynamic analysis for detecting memory leaks~\cite{b5}.
When memory is freed, a magic value is written into the memory. 
When the program exits, the heap is checked for memory leaks, i.e., by identifying allocated memory that has not been overwritten with a magic value.
Nethercote et al. in their framework Valgrind \cite{b11} presents a virtual architecture that captures all calls to memory allocation and deallocation by the program.
When the program exits, Valgrind checks whether a memory allocation has been freed or not.
If the allocation is not freed and it is also not reachable from stack and global variables, then it is considered to be a leak.
The approach suffers from performance overhead due to the synthetic execution, i.e., checking every memory access.

% The accuracy of dynamic based frameworks is dependant upon test coverage of the test suite, implying that it can miss out on several bugs in source program. However, these approaches have generally no false positives.

\vspace{1mm}
\noindent
Our proposed approach, AddressWatcher, suggests memory leak fixes. 
and is different from static and dynamic approaches for \textit{detecting} memory leaks.

% \iffalse
% AddressSanitizer (ASAN) \textbf{\cite{b2}} detects memory corruption vulnerabilities including stack and heap buffer overflows with the help of shadow memory and compile time instrumentation. However this approach also faces limitations. Firstly, it  cannot detect non contiguous memory violations (overflows that do not corrupt adjacent memory). Furthermore it cannot detect buffer overflows inside an object or allocated struct. AddressWatcher focuses only on fixing memory leaks and leverages ASAN infrastructure to achieve the goal of \mhmd{Can you add something here?}.
% \fi

\subsection{Approaches for Memory Leak Fixing}

\noindent
\textbf{Static approaches.}
\textcolor{black}{Some prominent techniques have been proposed to statically
suggest fixes for memory leaks~\cite{b13,b18}. 
%Several techniques like Memfix~\cite{b13} and Leakfix~\cite{b18} have been proposed 
For example, Memfix by Lee et al.~\cite{b13} is a static approach to fixing memory deallocation errors, including memory leaks.
First, Memfix identifies all program paths of a leaked object. Then, the bug fix is modeled as an Exact Cover problem where minimum frees must be placed to plug all the leaked paths. A solution to the exact cover problem is generated using a SAT solver. Then, all generated deallocation statements are inserted along the program paths.
However, Memfix is not precise and cannot track program paths in the presence of function pointers and recursion. Memfix often errors out when the number of program paths explodes and the SAT solver is unable to generate a solution.}

\textcolor{black}{Another static-based approach to fixing memory leaks is Leakfix~\cite{b18}. 
Leakfix performs pointer analysis on the whole program. Each procedure is classified into 3 types: those that allocate, deallocate, or use a given memory allocation. It then abstracts the program into a Control Flow Graph (CFG) where every node is a procedure classified into above three types. Then, the task of finding the correct deallocation is equivalent to finding edges in the graph that meet a set of criteria. These are program points where allocated memory is still in scope, but will never be used thereafter in the graph.
The precision of the approach is limited to the efficacy of pointer analysis techniques used for the classification of procedures (e.g., DSA~\cite{b43}.) Such techniques are employed to create the graph, but they face limitations. For example, when a pointer array contains two different allocated pointers, DSA cannot distinguish between them.}
%~\cite{b18}.

%\mhmd{If memfix is more recent than Leakfix, we should higlight that to justify why we only compared with it and not with leakfix.}
\textcolor{black}{Our proposed approach, AddressWatcher,
is complementary to such static-based analysis approaches.
For example, we compare AddressWatcher with Memfix using a dataset of 50 memory leak bugs. 
We find that AddressWatcher fixes 23 bugs related to non-error paths where Memfix errors out.}

\vspace{2mm}
\noindent
\textbf{Dynamic approaches.}
\textcolor{black}{Several dynamic-based techniques have been proposed to suggest different forms of leak fixes~\cite{b45,b12,b52}.
Yu et al., in their tool DEF\_LEAK, propose a dynamic symbolic execution approach to expose memory leaks occurring in all execution paths~\cite{b45}.
In their approach, the program to be analyzed is instrumented before execution. Dynamic symbolic execution is a technique of analyzing the program to determine inputs that would cause certain parts of the program to execute. This technique is employed to discover as many execution paths as possible. 
% During execution, pertinent information about allocated memory is delivered to a backend leak checker. The backend leak checker tracks the number of variables pointing to this allocated memory at all times. When there are no variables pointing to a given allocated memory, the leak checker automatically deallocates the memory. 
The approach suffers from inaccuracy when it encounters large programs whose inputs are hard to be symbolized in dynamic symbolic execution. AddressWatcher, on the other hand, relies on predefined test cases to suggest fixes, and hence, the size of the program or nature of input will not affect the accuracy of the suggested solution.
Instead, AddressWatcher relies on the completeness of the test suite. }
% Autofix~\cite{b19} is a static and dynamic analysis tool that uses static value flow analysis from allocated memory to identify potential locations where memory can be freed. It then instruments code in these locations where memory is automatically freed. The instrumentation in Autofix is added based on static value flow analysis in Saber~\cite{b1}, which is neither sound nor complete- bounding loops and recursion to utmost one iteration and does not capture path correlations. AddressWatcher adds instrumentation wherever heap variables are read/written and does not suffer from an imprecise value flow analysis.}
% LeakChaser~\cite{b52} is another dynamic analysis tool for Java and C# that allows programmers to assert object lifetime relationships (e.g., that one object dies before another). This approach requires developers’ knowledge which is essential for a leak detector to produce highly relevant reports. AddressWatcher on the other hand focuses on C/C++ programs for memory leak fixing and does not require programmer to stipulate object relationships.

\textcolor{black}{The work most close to ours is proposed by Clause et al. where they proposed LeakPoint ~\cite{b12}.
LeakPoint is a dynamic analysis framework that performs taint propagation on pointers to detect leaked objects, in order to identify last-use sites of the objects and suggest candidate sites for fixing them.
In fact, Leakpoint uses Valgrind infrastructure for implementing taint propagation and dynamic binary instrumentation, which brings a performance overhead of 100-300 times the base program~\cite{b12}.
% \textcolor{red}{you mean 100\% -- 300\% right, 100x is too much...}.
% AddressWatcher would have a better performance than Leakpoint, as AddressWatcher utilizes performance-optimized infrastructure of sanitizers such as ASAN and LSAN. 
% ASAN performance overhead can be a maximum of 3 times base program~\cite{b2}, while LSAN adds almost no performance overhead~\cite{b44}. 
We showed in Section~\ref{sec:eval} that AddressWatcher has a performance overhead of 2.42-2.78 times the base program.
 Unfortunately, we were not able to find a public implementation of all these dynamic tools, such as LeakPoint or DEF\_LEAK, which makes it infeasible to directly compare to AddressWatcher. AddressWatcher, on the other hand, is open source with a publicly available docker~\cite{b25}.}

% \subsection{Garbage Collection}
% The most widely used dynamic approaches to memory leak fixing is garbage collection \cite{b6,b7,b8,b9}. However, this is very hard in languages like C/C++ which do not have a clear distinction between pointers and data. This can lead to hidden pointers through type casting from pointers to integers, preventing garbage collectors from knowing when an object is truly not referenced by any pointer. The object can be freed only when a garbage collector knows that the object is not referenced by all pointers and data.
% This in principle leads to significant performance overhead. 
% AddressWatcher in comparison does not automatically free leaked memory at run time but only suggests to developers possible free locations for the fix.

\section{Conclusion and Future Work}
In this work,  we present AddressWatcher, a new framework for fixing memory leak bugs.
Previous static analysis approaches attempt to trace the complete semantics of memory objects on all leaked paths with imprecision. On the other hand, dynamic approaches attempt to profile the memory object on a particular execution path. AddressWatcher's novelty lies in the fact that it is a dynamic approach that allows the semantics of a memory object to be tracked over multiple execution paths using a leak database. 
Our evaluation reveals that AddressWatcher correctly suggests fixes for 23 out of 50 memory leaks in five benchmark open-source projects.
% Our further analysis indicates that AddressWatcher can be considered complementary to existing approaches to memory leak fixing.
% Compared to static-based approaches (i.e., Memfix), AddressWatcher can fix 12 memory leaks that Memfix cannot fix.
\textcolor{black}{Finally, we demonstrate AddressWatcher's practicality and relevance to the open-source community by submitting 25 PRs to 12 popular open-source repositories to fix memory leaks.
Out of these submissions, 21 PRs have been successfully approved and incorporated within the respective codebases. Furthermore, one of our fixes was deemed so critical that it prompted a new version release for the calc repository.}

 \def\UrlBreaks{\do\/\do-}

\bibliographystyle{IEEEtran}
\bibliography{Bibliography}

\vskip -2\baselineskip plus -1 fil
\begin{IEEEbiography}
[{\includegraphics[width=1.5in,height=1in,clip,keepaspectratio]{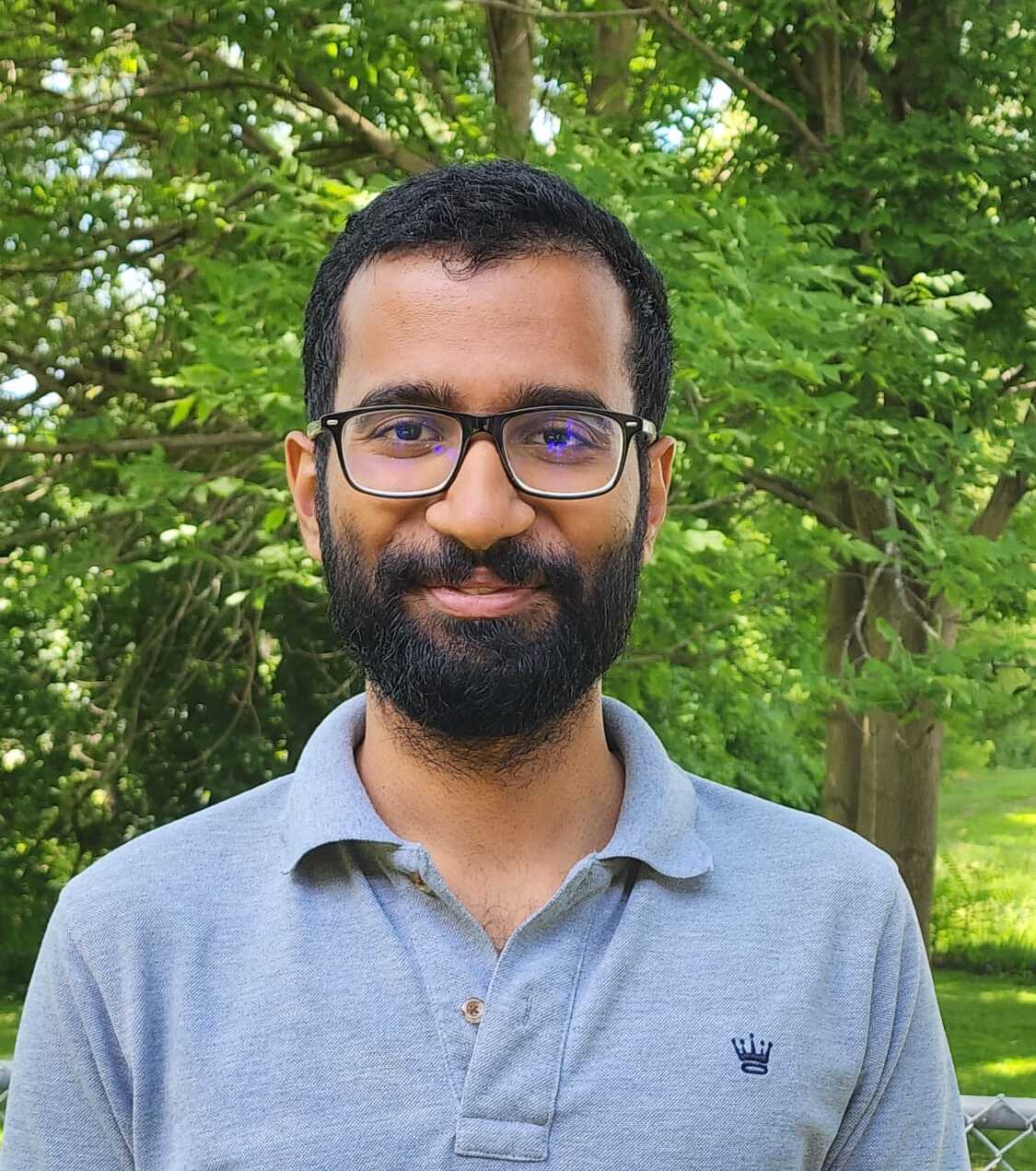}}]
{Aniruddhan Murali} is a Ph.D. candidate in the
Cheriton School of Computer Science at the
University of Waterloo, Canada. His research
interests include fuzzing, code slicing, vulnerability detection, and automatic bug fixing. You can find more about him \href{https://www.linkedin.com/in/aniruddhan-murali-40b41a11a/}{here}.
\end{IEEEbiography}

\vskip -4\baselineskip plus -1 fil
\begin{IEEEbiography}
[{\includegraphics[width=1.5in,height=1in,clip,keepaspectratio]{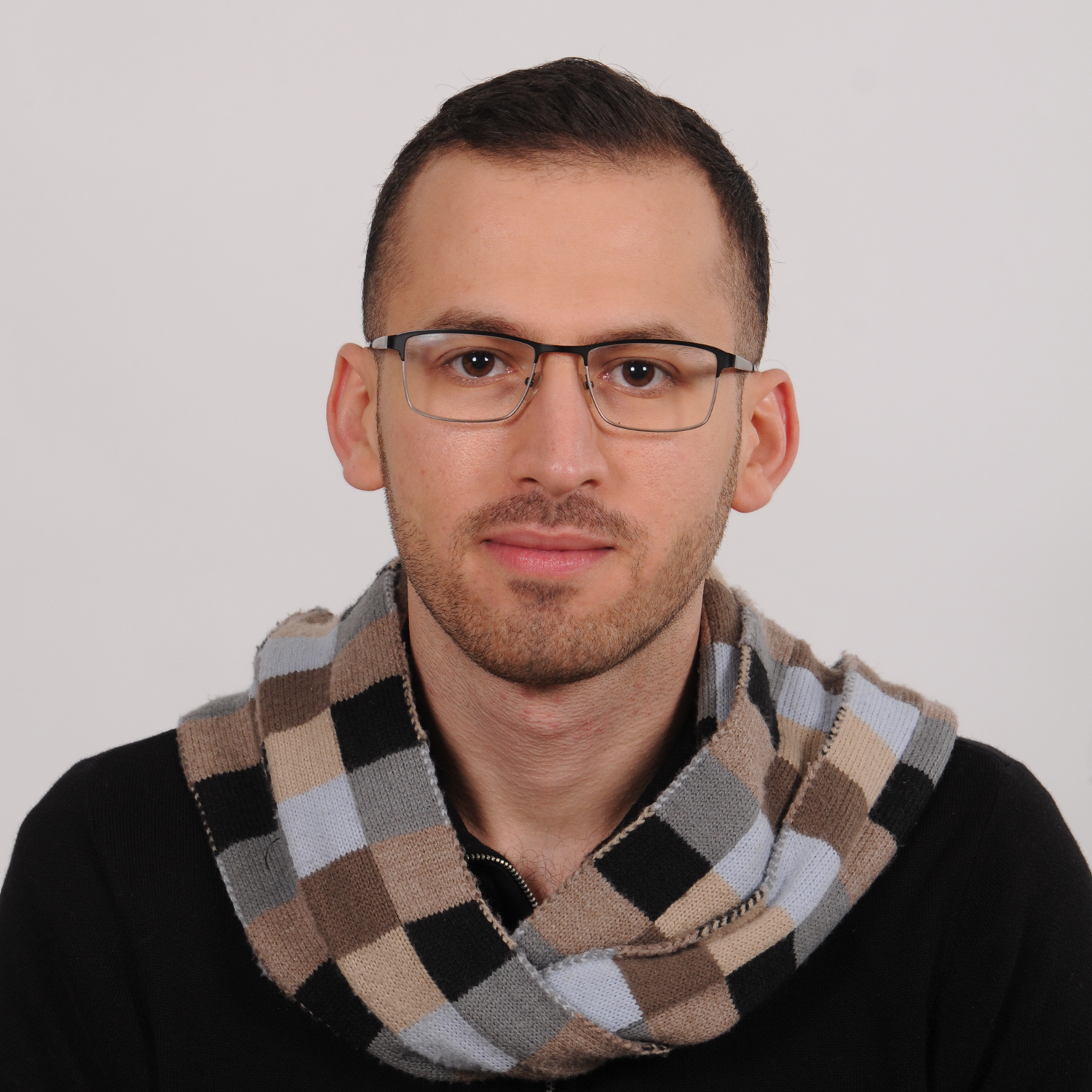}}]{Mahmoud Alfadel}  is a postdoctoral researcher
in the Cheriton School of Computer Science, University of Waterloo. His research interests include mining software repositories, empirical software engineering, software ecosystems, open-source security, and release engineering. You can find
more about him at \url{https://rebels.cs.uwaterloo.ca/member/mahmoud.html}
\end{IEEEbiography}

\vskip -4\baselineskip plus -1fil
\begin{IEEEbiography}
[{\includegraphics[width=1.5in,height=1in,clip,keepaspectratio]{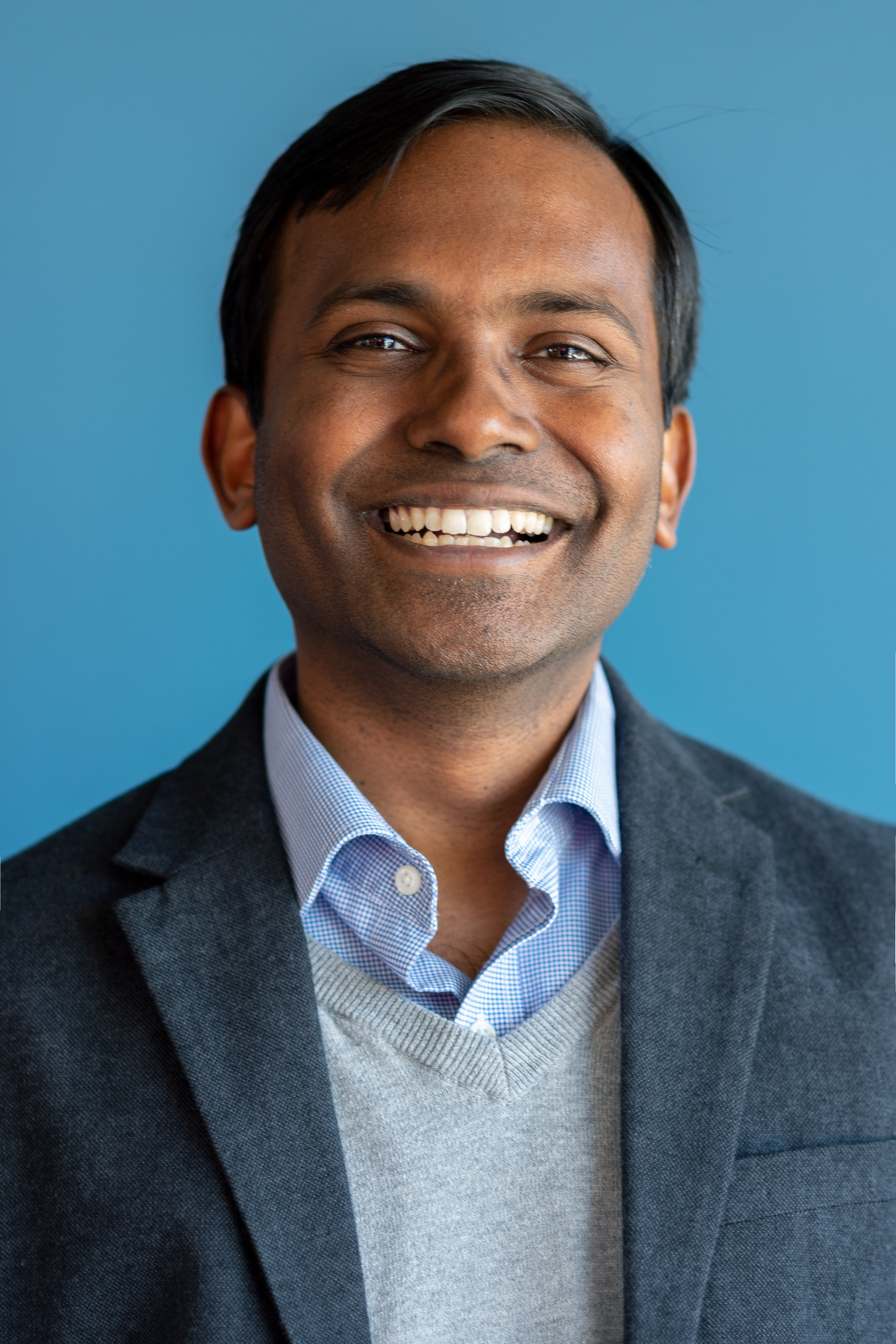}}]{Meiyappan Nagappan} is an Associate Professor at the Cheriton School of Computer Science, University of Waterloo. He has worked on empirical software engineering to address software development concerns and currently researches the impact of large language models on software development.
\end{IEEEbiography}

\vskip -4\baselineskip plus -1fil
\begin{IEEEbiography}
[{\includegraphics[width=1.5in,height=1in,clip,keepaspectratio]{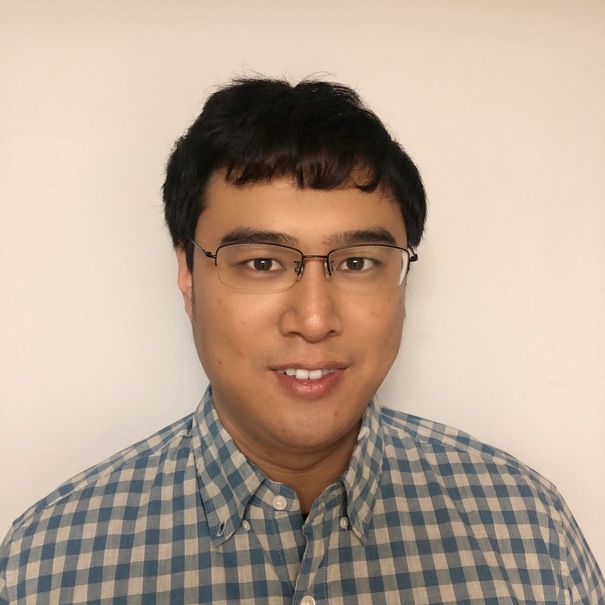}}]{Meng Xu} is an Assistant Professor in the Cheriton School of
Computer Science at the University of Waterloo, Canada. His research is
in the area of system and software security, with a focus on delivering
high-quality solutions to practical security programs, especially in
finding and patching vulnerabilities in critical computer systems. This
usually includes research and development of automated program analysis
/ testing / verification tools that facilitate the security reasoning
of critical programs.
\end{IEEEbiography}

\vskip -2\baselineskip plus -1fil
\begin{IEEEbiography}
[{\includegraphics[width=1.5in,height=1in,clip,keepaspectratio]{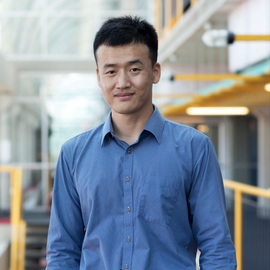}}]{Chengnian Sun} is currently an Associate Professor at the Cheriton School of Computer Science, University of Waterloo, Canada. His research interests include software engineering and programming languages, with a focus on techniques, tools, and methodologies for improving software quality and developer productivity. He received his Ph.D. degree from the School of Computing at the National University of Singapore. 
\end{IEEEbiography}

%====
\end{document}